\documentclass[useAMS,usenatbib]{mn2e}
\pdfoutput=1 
\usepackage{hyperref}
\usepackage{color}
\usepackage{amsmath}
\usepackage{graphicx}
\usepackage{txfonts}
\usepackage{natbib}
\usepackage{pdflscape} 

\newcommand{\teff}{T_{\rm{eff}}}

\newcommand{\logg}{\log(g)}
\newcommand{\feh}{\textrm{[Fe/H]}}

\newcommand\lta{\mathrel{\hbox{\raise 0.6 ex \hbox{$<$}\kern
                   -1.8 ex\lower .5 ex\hbox{$\sim$}}}}
\newcommand\gta{\mathrel{\hbox{\raise 0.6 ex \hbox{$>$}\kern
                   -1.7 ex\lower .5 ex\hbox{$\sim$}}}}
\newcommand{\RN}[1]{{\small\textup{\uppercase\expandafter{\romannumeral#1}}}}

\title[Colour$-\teff$ relations in the Gaia system]{The GALAH survey: effective temperature calibration from the InfraRed Flux Method in the Gaia system}

\author[Casagrande et al.]{
  Luca~Casagrande$^{1,2}$\thanks{E-mail: \href{mailto:luca.casagrande@anu.edu.au}{luca.casagrande@anu.edu.au}}\thanks{Colour-$\teff$ routines: \href{https://github.com/casaluca/colte}{\textcolor{blue}{https://github.com/casaluca/colte}}},
Jane~Lin$^{1,2}$,
Adam~D.~Rains$^{1}$,
Fan~Liu$^{3}$,
Sven~Buder$^{1,2}$,\newauthor
Jonathan~Horner$^{4}$,
Martin~Asplund$^{5}$,
Geraint~F.~Lewis$^{6}$,
Sarah~L.~Martell$^{7,2}$,\newauthor
Thomas~Nordlander$^{1,2}$,
Dennis~Stello$^{7,2}$,
Yuan-Sen~Ting$^{8,9,10,1,2}$,
Robert~A.~Wittenmyer$^{4}$,\newauthor
Joss~Bland-Hawthorn$^{6,2}$,
Andrew~R.~Casey$^{11,12}$,
Gayandhi~M.~De~Silva$^{13,14}$,\newauthor
Valentina~{D'Orazi}$^{15}$,
Ken~C.~Freeman$^{1}$,
Michael~R.~Hayden$^{6,2}$,
Janez~Kos$^{16}$,
Karin~Lind$^{17}$,\newauthor
Katharine~J.~Schlesinger$^{1}$,
Sanjib~Sharma$^{6,2}$,
Jeffrey~D.~Simpson$^{7,2}$,
Daniel~B.~Zucker$^{18,14}$,\newauthor
Toma\v{z}~Zwitter$^{16}$
\\
\\
(Affiliations listed after the references)}

\begin{document}

\date{Received; accepted}

\maketitle

\begin{abstract}
  In order to accurately determine stellar properties, knowledge of the
  effective temperature of stars is vital. We implement Gaia and 2MASS
  photometry in the InfraRed Flux Method and apply it to over 360,000 stars
  across different
  evolutionary stages in the GALAH DR3 survey. We derive colour-effective
  temperature relations that take into account the effect of metallicity and
  surface gravity over the range $4000\,\rm{K}\lesssim\teff\lesssim8000\,\rm{K}$, from very metal-poor stars to super solar metallicities.
  The internal uncertainty of these calibrations is of order 40$-$80 K
  depending on the colour combination used. Comparison against solar-twins,
  Gaia benchmark stars
  and the latest interferometric measurements validates the precision and
  accuracy of these calibrations from F to early M spectral types. We assess the
  impact of various sources of uncertainties, including the assumed
  extinction law, and provide guidelines to use our relations. Robust
  solar colours are also derived.
\end{abstract}


\section{Introduction}

The effective temperature ($\teff$) is one of the most fundamental stellar
parameter, and it affects virtually every stellar property that we determine,
be it from spectroscopy, or inferred by comparing against stellar models
\citep[e.g.,][]{ng18,choi18}. 

While angular diameters measured from interferometry provide the most direct way
to measure effective temperatures of stars \citep[provided bolometric fluxes
  can also be determined, see e.g.,][]{code76}, they require a considerable
investment of time. Such analysis require a careful assessment of systematic
uncertainties, and they are biased towards bright targets, which are often
saturated in modern photometric systems and all-sky surveys
\citep[e.g.,][]{w13,lachaume19,rains20}. Further, these stars are often the
hardest to observe for large-scale spectroscopic surveys. 

Among the many indirect methods to determine $\teff$ is the InfraRed Flux Method
(hereafter IRFM), an almost model independent photometric technique originally
devised to obtain angular diameters to a precision of a few per cent, and
capable of competing against intensity interferometry in cases where a good flux
calibration is achieved \citep[][]{bs77,b79,b80}. Over the years, the IRFM has
been successfully applied to determine the effective temperatures of stars of
different spectral types and metallicities
\citep[e.g.,][]{bs77,al96,al99,rm05,ghb09,c10}.

The version of the IRFM used in this work has been previously validated against
solar twins, HST absolute spectrophotometry and interferometric angular
diameters \citep{c06,c10}. In particular, dedicated near-infrared photometry has
been carried out to derive effective temperatures of interferometric targets
with saturated 2MASS magnitudes \citep{c14}. Our $\teff$ scale is widely used
by many studies and surveys, and we now make it available into the Gaia
photometric system. To do so, we implement Gaia photometry into the IRFM
described in \cite{c06,c10}. Also, thanks to Gaia parallaxes it is now possible
to derive reliable surface gravities. We provide
colour$-\teff$ relations which take into account the effect of metallicity and
surface gravity by running the IRFM for all stars in the third Data Release
(DR3)
of the GALAH survey \citep{buder20}. This data release also includes stars
observed with the same instrument setup, data reduction and analysis pipeline
by the K2-HERMES \citep{k2-hermes1,k2-hermes2} and TESS-HERMES
\citep{tess-hermes} surveys.

We describe how Gaia photometry is implemented into our version of the IRFM in
Section \ref{sec:irfm} and present colour$-\teff$ relations in Section
\ref{sec:cal}. We benchmark our results against standard stars, assess the
typical $\teff$ uncertainty of our calibrations and provide guidelines for their
use in Section \ref{sec:test}. Finally, we comment on the use of different
colour indices and draw our conclusions in Section \ref{sec:finale}.

\section{The InfraRed Flux Method using Gaia photometry}\label{sec:irfm}

The IRFM can be viewed as the most extreme colour technique since it relies
on the
index defined by the ratio between the bolometric and the infrared
monochromatic flux of a star. This ratio can be compared to that obtained
using the same quantities defined on a stellar surface element, $\sigma \teff^4$
and $\mathcal{F}_\mathrm{IR}\rm{(model)}$, respectively \citep[see e.g.,][]{al96bis,c06}.
If stellar and model fluxes are known, it is then possible to solve for $\teff$.
As we describe
later, this step is done iteratively in our version of the IRFM. The
crucial advantage of the IRFM over other colour techniques is that, at
least for
spectral types hotter than early M-type, near-infrared photometry samples the 
Rayleigh--Jeans tail of stellar spectra, a region largely dominated by the
continuum\footnote{See however \citealt{black91} for a discussion of the
  importance of $\rm{H}^{-}$ opacity.}, with a roughly linear dependence on
$\teff$.
The model dependent term  $\mathcal{F}_\mathrm{IR}\rm{(model)}$ is almost
unaffected by metallicity, surface gravity and granulation, as extensively
tested in the literature \citep[e.g.,][]{al96,agp01,rm05,c06,c09,ghb09}.

We use the implementation of the IRFM described in \cite{c06,c10}, where for
each star we now use Gaia $BP,RP$ and 2MASS $JHK_s$ photometry to derive the
bolometric flux. The flux outside these bands (i.e., the bolometric correction)
is estimated using a theoretical model flux at a given $\teff, \logg$ and
[Fe/H]. The infrared monochromatic flux is derived from 2MASS magnitudes only.
An iterative procedure in $\teff$ is adopted to cope with the mildly
model-dependent nature of the bolometric correction and surface infrared
monochromatic flux. We interpolate over the \cite{ck03} grid of model
fluxes, starting for each star with an initial estimate of its effective
temperature and adopting the GALAH DR3 [Fe/H] and $\logg$, until convergence
in $\teff$ is reached within 1K. The convergence is robust regardless of
  the initial $\teff$ estimate. The model dependence is expected to be small,
  and in \cite{c06,c10} we tested that using the MARCS grid of model fluxes
  \citep{gustafsson08} affects the resulting $\teff$ only by few K for dwarfs
  and subgiants in the range $\simeq 4500-6500$~K.

For Gaia $BP$ and $RP$ magnitudes we use the {\tt Gaia-DR2} formalism described
in \cite{cv18}, which is based on the revised transmission curves and
non-revised Vega zero-points provided by \cite{evans18}. As described in
\cite{cv18}, this choice best mimics the photometric processing done by the
Gaia team to reproduce {\tt phot\_g\_mean\_mag}, {\tt phot\_bp\_mean\_mag}
and {\tt phot\_rp\_mean\_mag} given in Gaia DR2. In Appendix \ref{appA} we also
implement Gaia EDR3 photometry, and provide calibrations for this system. We
remark that although Gaia EDR3 is formally an independent photometric systems
from Gaia DR2, differences are overall small for the sake of the $\teff$
derived from the IRFM (although the calibrations in the two systems should not
be used interchangeably, as further discussed in Appendix \ref{appA}).
We use $BP$ and $RP$ instead of $G$ magnitudes for a number of reasons:
comparison with absolute spectrophotometry indicates that $BP$ and $RP$
are reliable and well standardized in the magnitude range $\simeq$ 5 to 16,
which is relevant for our targets. On the contrary, $G$ magnitudes have a
magnitude dependent offset, and are affected by uncalibrated CCD saturation for
$G\lesssim6$ \citep{evans18,cv18,maw18}. Further, the $BP$ and $RP$ bandpasses
together have the same wavelength coverage as the $G$ bandpass.
\begin{figure}
\begin{center}
\includegraphics[width=0.45\textwidth]{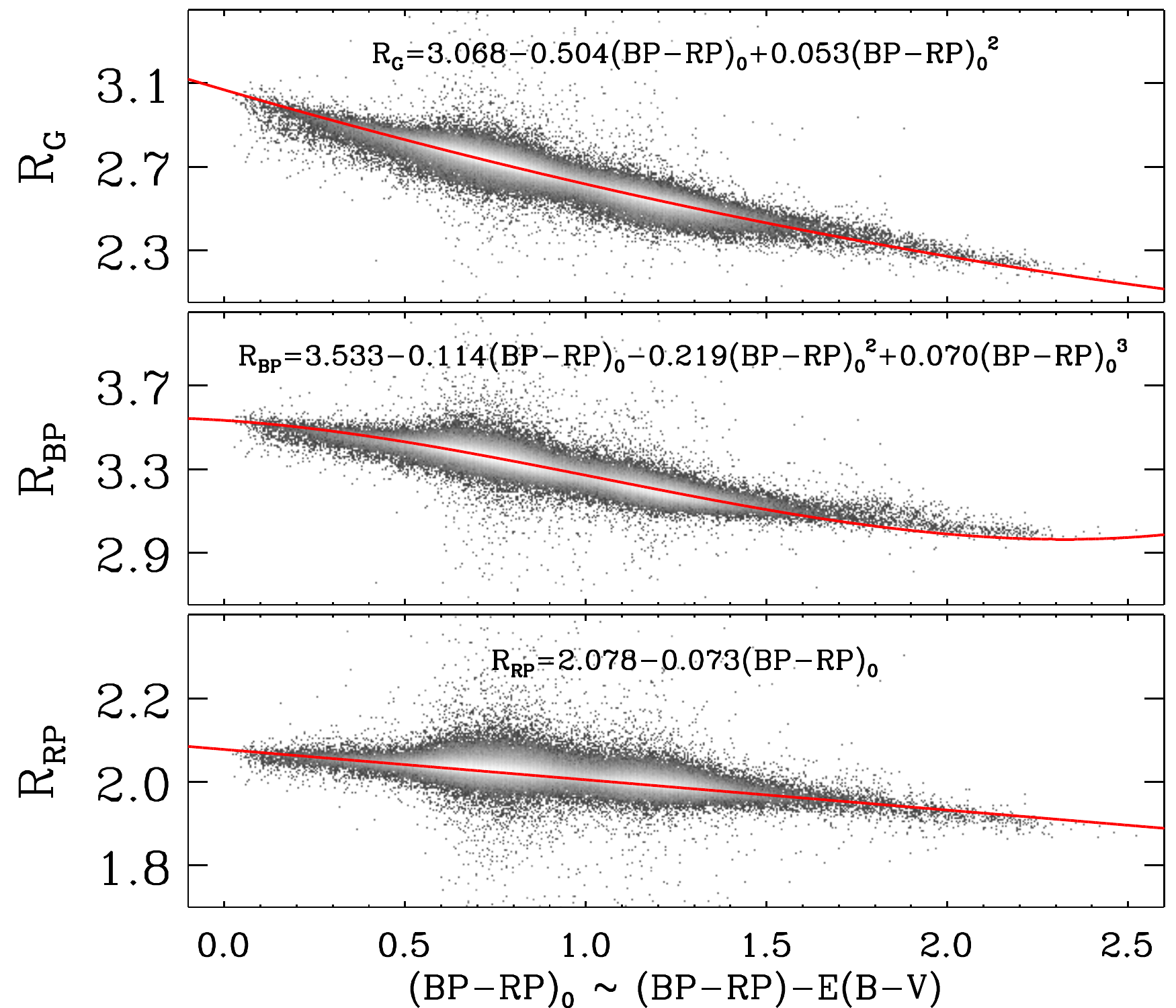}
\caption{Gaia extinction coefficients as function of intrinsic stellar
    colours for our sample of stars (colour coded in gray by log-density). Red
    solid lines
    show the fits given in each panel. To estimate intrinsic colours needed for
    the fits, one can iterate starting from
  $(BP-RP)_0 \simeq (BP-RP)-E(B-V)$. See Appendix
  \ref{appB} for a summary of the extinction coefficients for Gaia DR2, EDR3
  and 2MASS under different extinction laws.}\label{fig:exco} 
\end{center}
\end{figure}

One of the most critical points when implementing the IRFM is the photometric
absolute calibration (i.e., how magnitudes are converted into fluxes), which
sets the zero-point of the $\teff$ scale. This is particularly important in
the infrared, for which we use the same 2MASS prescriptions discussed in
\cite{c10}. To verify that the zero-point of our $\teff$ scale is not altered
by Gaia magnitudes, we derive $\teff$ for all stars in \cite{c10} with a
counterpart in Gaia (408 targets). Not unexpectedly, we find excellent
agreement, with both mean and median $\Delta\teff=12 \pm 2$~K
($\sigma=41$~K) and no trends as function of stellar parameters. This
difference is robust, regardless of whether the stars used are those with the
best Gaia quality flags. Although this
difference is fully within the 20~K zero-point uncertainty of the reference
$\teff$ scale of \cite{c10}, we correct for this small offset to adhere to the
parent scale.

We apply the IRFM to over 620,000 spectra in GALAH DR3 for which [Fe/H], $\logg$,
$BP,RP,J,H,K_s$ are available. About 40 percent of the targets have $E(B-V)$
from \cite{bayestar19}. For the remaining stars, we rescale reddening from
\cite{sfd} with the same procedure described in \cite{c19}.
Effective temperatures from the IRFM along with adopted values of reddening
are available as part of GALAH DR3 \citep[][which also includes a comparison
against the GALAH spectroscopic $\teff$.]{buder20}. To account for the
spectral type dependence of extinction coefficients, in the IRFM we adopt the
\cite{ccm89}/\cite{od94} extinction law, and for each star compute extinction
coefficients with the synthetic spectrum at the $\teff, \logg$ and [Fe/H] used
at each iteration. 

Figure \ref{fig:exco} shows extinction coefficients
for the Gaia filters as function of intrinsic (i.e.,~reddening corrected)
stellar colour for our sample of stars. For the 2MASS system there is
virtually no dependence on
spectral type and the following constant values are found $R_J=0.899$,
$R_H=0.567$ and $R_{Ks}=0.366$. These coefficients are in excellent agreement
with those reported in \cite{cv14,cv18}, obtained using the same extinction
law. We discuss in Appendix \ref{appB} the effect of using different
extinction laws on the derived colour$-\teff$ relations and extinction
coefficients.

\begin{figure}
\begin{center}
\includegraphics[width=0.48\textwidth]{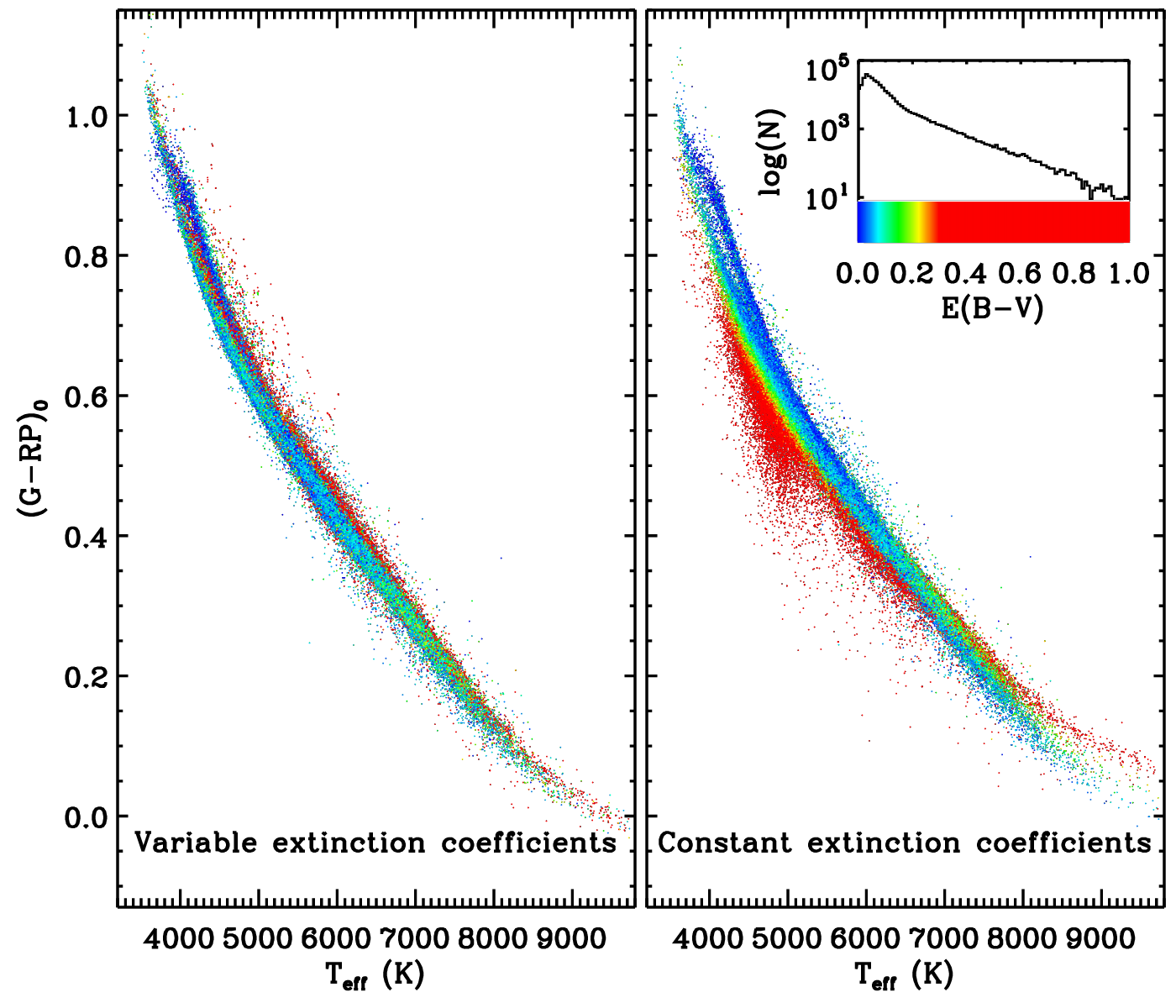}
\caption{Left panel: colour-$\teff$ relation obtained from the IRFM in
  $(G-RP)_0$, where extinction coefficients are computed for each star
  individually. Right
  panel: colour-$\teff$ relation using the same input effective temperatures, 
  but constant extinction coefficients to deredden the colour index. The
  importance of using variable extinction coefficients becomes visible for
  increasing values of reddening. Stars are colour coded by their $E(B-V)$ with
  the distribution shown in the inset.}\label{fig:compa} 
\end{center}
\end{figure}

The use of constant extinction coefficients instead of colour dependent ones
affects colour indices, and hence the effective temperatures derived from the
relations of Section \ref{sec:cal}. This can be
appreciated from the comparison in Figure \ref{fig:compa}, where the difference
in colour obtained using constant or colour dependent extinction coefficients
is amplified at high values of reddening for a given input $\teff$ . The fits
of Figure
\ref{fig:exco} should thus be preferred to deredden colour indices involving
Gaia bands, especially in regions of high extinction. 

\section{Colour$-\teff$ relations}\label{sec:cal}

In order to derive colour-$\teff$ relations, we first apply a few quality cuts.
We restrict ourselves to stars with the best GALAH DR3 spectroscopic
parameters ({\tt flag\_sp=0}), and Gaia photometry $1.0+0.015\,(BP-RP)^2<$ {\tt phot\_bp\_rp\_excess\_factor} $<1.3+0.060\,(BP-RP)^2$ and
{\tt phot\_proc\_mode=0}. There is a sharp drop in the number of stars
  with $G>14$, and this reflects the GALAH selection function. Only 5 percent
  of stars are fainter than 14, and $0.06$ percent are in the faintest bin
  $16 < G \lesssim 16.5$. For relations involving the $G$ band we also exclude
a handful of stars with $G<6$ \citep{evans18,riello18}. These requirements
yield automatically good 2MASS photometry: median photometric errors in $JHK_s$
are $0.024$ mag with $99.9$ percent of the targets having 2MASS quality flag
{\tt Qflg='AAA'}.
\begin{figure*}
\begin{center}
\includegraphics[width=1\textwidth]{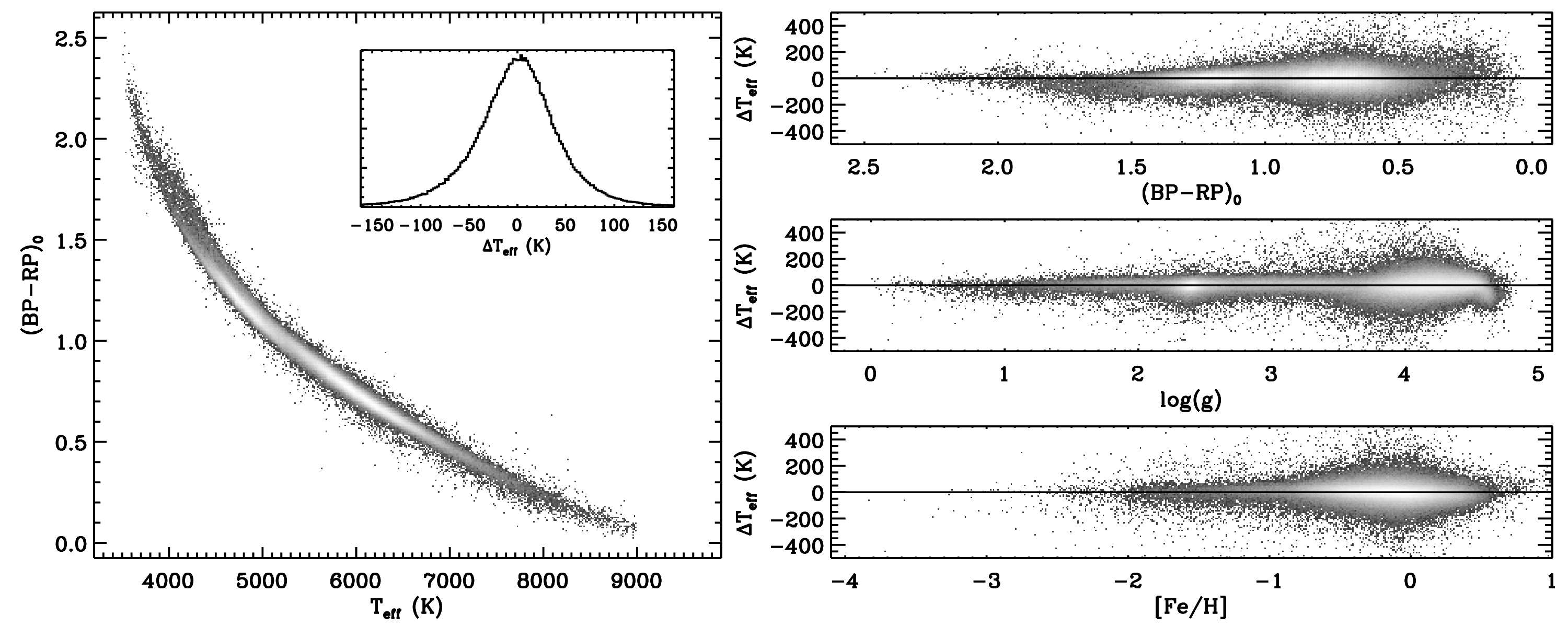}
\caption{Left panel: log-density plot of the colour-$\teff$ relation obtained
  using all 360,000 GALAH DR3 stars with good photometric and spectroscopic
  flags as described in the text. For $\teff \lesssim 4500$~K the two loci
  defined by dwarf and giant stars can be noticed. The inset shows the
  distribution of the $\teff$ residuals of our calibration. Right panels:
  $\teff$ residuals plotted as function of colour, surface gravity and
  metallicity. Plots for the other colour indices are available as
  supplementary online material.}\label{fig:bprp} 
\end{center}
\end{figure*}

Depending on the combination of filters, there are over 360,000 stars
available for our fits. We use only stars with $E(B-V)<0.1$ to derive our
  fits, to avoid a strong dependence on the adopted extinction law (Appendix
  \ref{appB}). Due to the combined effect of the GALAH selection
function and target selection effects (most notably stellar evolutionary
timescales), the distribution of targets has two main temperature overdensities:
one at the main-sequence turn-off and the other at the red-clump phase. If all
available stars were used to derive colour$-\teff$ relations these two
overdensities would dominate the fit. Instead, we sample our stars uniformly in
$\teff$, randomly selecting 20 stars every 20~K, and repeating this for 10
realizations. The calibration sample for each fit is thus based on roughly
50,000 stars. We repeat the above procedure 10,000 times, and select the fit
that returns the lowest standard deviation with respect to the input effective
temperatures from the IRFM. We also explored the effect of a uniform gridding
in $\teff$ and $\logg$ but did not find any significant difference with
respect to a uniform sampling in $\teff$ only.

\begin{figure*}
\begin{center}
\includegraphics[width=1\textwidth]{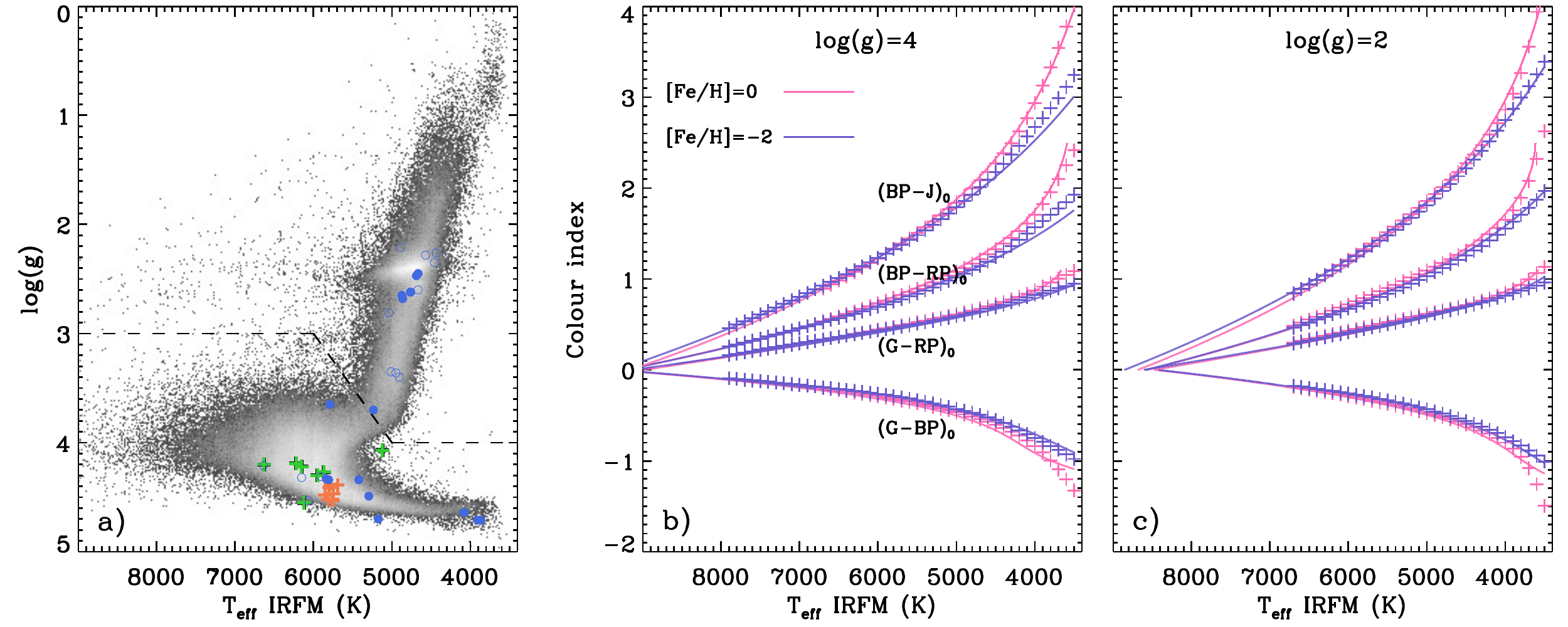}
\caption{Panel a): Kiel diagram of the GALAH DR3 sample used to derive the
  colour-$\teff$ relations presented in this work. The dashed line marks the
  separation
  between dwarf and giant stars discussed in Section \ref{sec:test}. Coloured
  crosses and circles are the stars used in Fig \ref{fig:benc} to test the
  $\teff$ scale. Panel b) and c): some of the colour-$\teff$ relations (solid
  lines) of Table \ref{tab:fit} for
  fixed values of $\logg=4$ and $\logg=2$, and $\feh=0$ and $\feh=-2$, as
  labelled. Plotted for comparison are synthetic colour-$\teff$ computed for
  the same values of gravity and metallicity (cross symbols). Note that the
  maximum $\teff$ available for synthetic colours varies with the adopted
  $\logg$.}\label{fig:test} 
\end{center}
\end{figure*}

To derive our relations we started with a polynomial as a function of colour,
which is the parameter that has the strongest dependence on $\teff$. Depending
on the colour index, we found that a third or fifth order polynomial was
necessary to describe the curve inflection occurring at low $\teff$. We then
added the $\feh$ and $\logg$ dependence into the fit. The Gaia broad band
filters have a rather mild dependence on metallicity, and the effect of $\logg$
is most noticeable below 4500~K, where colour-$\teff$ relations for dwarf
and giant stars branch off (Figure \ref{fig:bprp} and \ref{fig:test}). We
found no need to go higher than first order in $\feh$ and $\logg$, but
cross-terms with colour, as well as a term involving colour, $\teff$ and
$\logg$ were found to improve the fit. The adopted functional
form is:
\begin{displaymath}
\teff = a_0 + a_1 X +   a_2 X^2 + a_3 X^3 + a_4 X^5 + a_5 \logg + a_6 \logg\,X + 
\end{displaymath}
\vspace{-0.5cm}
\begin{displaymath}
a_7 \logg\,X^2 + a_8 \logg\,X^3 + a_9 \logg\,X^5 + a_{10} \feh + 
\end{displaymath}  
\vspace{-0.5cm}
\begin{equation}\label{xx}
a_{11} \feh\,X + a_{12} \feh\,X^2 + a_{13} \feh\,X^3 + a_{14}\feh\logg\,X
\end{equation}  
where $X$ is the colour index corrected for reddening, and not all terms
were found to be significant for all colour indices. The coefficients of
Eq.\,\ref{xx} are given in Table \ref{tab:fit}. Our relations and associated
standard deviations are derived over the range
$3600\,\rm{K}\lesssim \teff \lesssim 9000$~K, although as we discuss in the
next Section, they are validated by independent measurements over a smaller
range of effective temperatures. Polynomial fits are also typically less robust
towards the edges of a colour index. In Table \ref{tab:range} we recommend
conservative colour ranges, which effectively limit the applicability of our
relations between 4000~K and 8000~K for most filter combinations. 

\setcounter{table}{1}
\begin{table}
\caption{Recommended colour range for the validity of our calibrations.}\label{tab:range}
\begin{tabular}{ccc}
\hline
colour & dwarfs & giants \\
\hline
$(BP-RP)_0$  &  $[\phantom{+}0.20,\phantom{+}2.00]$ & $[\phantom{+}0.20,\phantom{+}2.55]$  \\
$(G-BP)_0$   &  $[-1.00,-0.15]$                     & $[-1.40,-0.15]$                      \\ 
$(G-RP)_0$   &  $[\phantom{+}0.15,\phantom{+}0.85]$ & $[\phantom{+}0.15,\phantom{+}1.15]$  \\
$(BP-J)_0$   &  $[\phantom{+}0.25,\phantom{+}3.00]$ & $[\phantom{+}0.90,\phantom{+}4.20]$  \\
$(BP-H)_0$   &  $[\phantom{+}0.40,\phantom{+}4.00]$ & $[\phantom{+}0.40,\phantom{+}4.90]$  \\
$(BP-K_s)_0$ &  $[\phantom{+}0.30,\phantom{+}4.20]$ & $[\phantom{+}0.30,\phantom{+}5.30]$  \\
$(RP-J)_0$   &  $[\phantom{+}0.20,\phantom{+}1.05]$ & $[\phantom{+}0.60,\phantom{+}1.55]$  \\
$(RP-H)_0$   &  $[\phantom{+}0.20,\phantom{+}1.60]$ & $[\phantom{+}0.20,\phantom{+}2.45]$  \\
$(RP-K_s)_0$ &  $[\phantom{+}0.20,\phantom{+}1.85]$ & $[\phantom{+}0.20,\phantom{+}2.70]$  \\
$(G-J)_0$    &  $[\phantom{+}0.15,\phantom{+}2.10]$ & $[\phantom{+}1.00,\phantom{+}2.80]$  \\
$(G-H)_0$    &  $[\phantom{+}0.25,\phantom{+}2.60]$ & $[\phantom{+}0.25,\phantom{+}3.70]$  \\ 
$(G-K_s)_0$  &  $[\phantom{+}0.20,\phantom{+}2.80]$ & $[\phantom{+}0.20,\phantom{+}3.90]$  \\
\hline
\end{tabular}
\begin{minipage}{1.4\textwidth}
Dwarfs and giants are separated as per Figure \ref{fig:test}a.
\end{minipage}
\end{table}

Figure \ref{fig:bprp} shows the colour-$\teff$ relation for $(BP-RP)_0$, along
with the residuals of the fit as function of colour, gravity and metallicity. 
Although Eq.\,(\ref{xx}) virtually allows for any combination of input
parameters, it should be recalled that stars distribute across the HR diagram
as permitted by stellar evolutionary theory. Figure \ref{fig:test}a illustrates
the range of stars used to build our colour calibrations, where cool stars are
found both at low and high surface gravities, whilst the hottest stars have
$\logg\sim4$. Fig \ref{fig:test}b and \ref{fig:test}c show the dependence on
$\logg$ and $\feh$ for some of our colour-$\teff$ calibrations. In addition,
to allow direct comparison, we also plot predictions from synthetic stellar
fluxes computed with the
{\tt bolometric-corrections}\footnote{\href{https://github.com/casaluca/bolometric-corrections}{\textcolor{blue}{https://github.com/casaluca/bolometric-corrections}}}
code \citep{cv14,cv18}. The purpose of this comparison is not to validate
empirical nor theoretical relations, but to show that our functional form well
captures the expected change of colours with $\teff$, $\logg$ and $\feh$. Some
of the discrepancies between empirical and theoretical predictions at the
coolest $\teff$ are likely due to inadequacies of synthetic fluxes as
discussed in the literature \citep[see e.g.,][]{cv14,topcu}.

\section{Validation and uncertainties}\label{sec:test}

We validate our colour-$\teff$ relations using three different test
populations and approaches, focusing on Solar twins, Gaia Benchmark Stars
(GBS), and interferometric measurements. The stars used for this purpose are
some of the brightest and best observed in the sky, with careful
determinations of their stellar parameters. In all instances, we apply the same
requirements on {\tt phot\_bp\_rp\_excess\_factor} and {\tt phot\_proc\_mode}
discussed in Section \ref{sec:cal} to select the best photometry. We also
exclude stars with $G<6$ and $BP$ and $RP<5$ due to uncalibrated systematics
at bright magnitudes. We only use 2MASS photometry with {\tt Qflg='A'} in a
given band.

\begin{figure*}
\begin{center}
\includegraphics[width=1\textwidth]{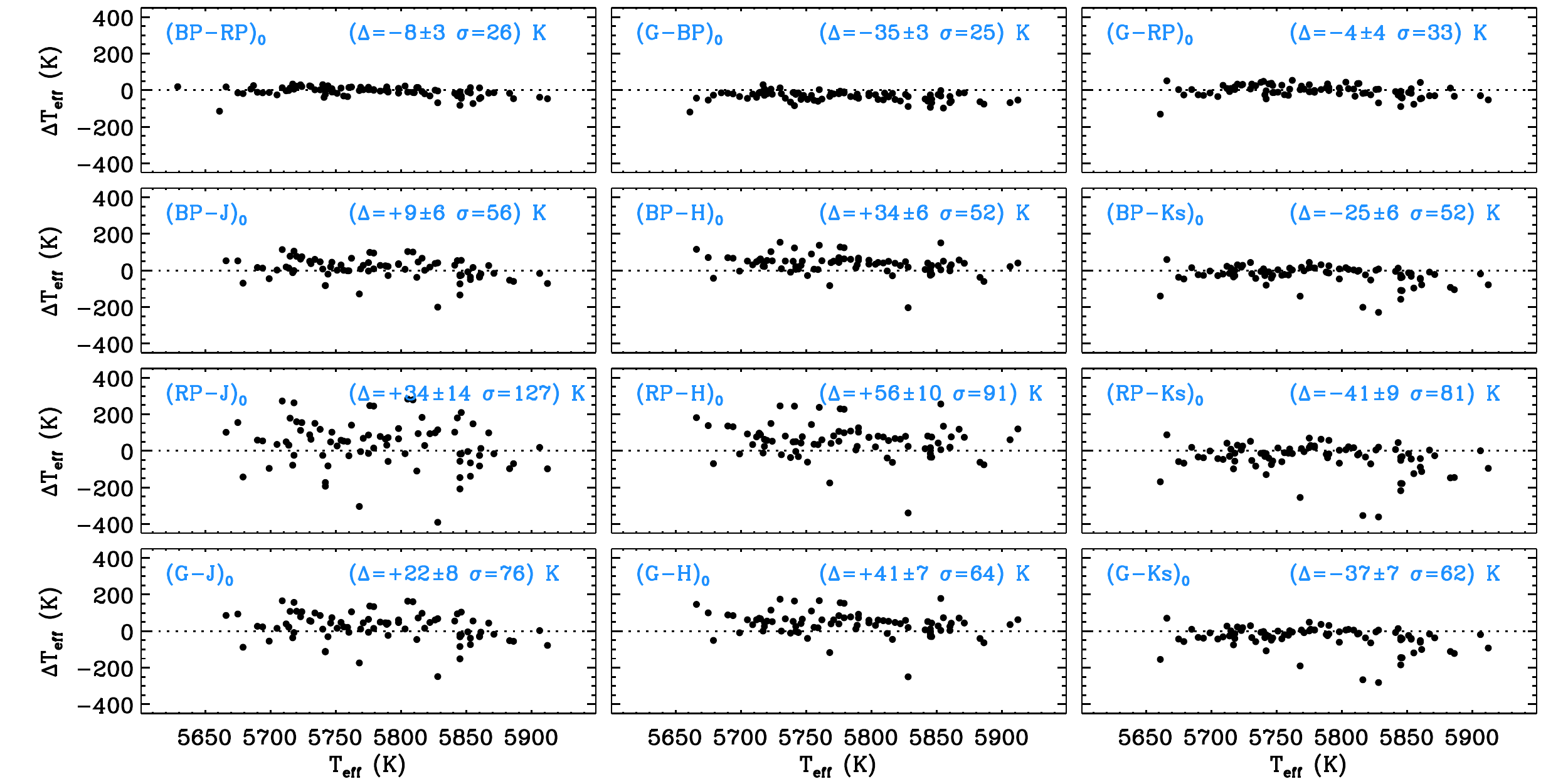}
\caption{Comparison between the effective temperatures obtained
  from our calibrations and those derived by Nissen (2015) and
  Spina et al.~(2018) from differential spectroscopic analysis of solar
  twins (x-axes). All targets are closer than 100pc and unaffected by reddening.
  In each panel we report the colour index used, the mean difference
  $\Delta(\rm{ours}-\rm{spectroscopy})$ $\pm$ standard error of the mean,
  and standard deviation ($\sigma$). Median and mean differences agree to
  within a few K.}\label{fig:nb} 
\end{center}
\end{figure*}

The sample of solar twins is the same that was used by \cite{c10} to set the
zero-point of their
$\teff$ scale. These twins are all nearby, unaffected by reddening, and with
good Gaia and 2MASS photometry. Accurate and precise spectroscopic $\teff$,
$\logg$ and $\feh$ are available from differential analysis of high-resolution,
high S/N spectra with respect to a solar reference spectrum, using
  excitation and ionization balance of iron lines \citep{mel06,melendez}. In
  particular, the identification of the best twins is based on the measured
  relative difference in equivalent widths and equivalent widths vs.
  excitation potential relations with respect to the observed solar reference
  spectrum, and thus entirely model independent.
In Table \ref{tab:check} we report the mean difference between the effective
temperatures
we derive in a given colour index, and the spectroscopic ones. Our $\teff$
are typically within few degrees of the spectroscopic ones. Further,
regardless of the spectroscopic effective temperatures, the mean and median
$\teff$ for our sample of solar twins in any colour index is always within few
tens of K of the solar $\teff$. The fact that our colour-$\teff$ relations are
well calibrated around the solar value is not unexpected, but confirms that we
have achieved our goal of tying the current $\teff$ scale to that of
\cite{c10}. To further test our scale, we use a large sample of more than 80
solar twins from \cite{Nissen15} and \cite{Spina18}. Also these twins have
highly accurate
and precise stellar parameters due to differential
spectroscopic analysis. This means that in the comparison we are essentially
dominated by photometric errors and intrinsic uncertainty in our colour-$\teff$
relations. The comparison in Figure \ref{fig:nb} shows that the standard
deviations for each colour index are consistent with the values reported in
Table \ref{tab:fit}, although the latter are derived over a much larger range of
stellar parameters. For solar type stars $(BP-RP)_0$, $(G-BP)_0$ and $(G-RP)_0$
are the colours with the highest precision, whilst the use of $RP$
photometry with 2MASS is the least informative, as it carries a typical
uncertainty of order $100$~K. The standard error of the mean shows that
individual colours can have systematic offsets of a few tens of K at most:
although calibrations are built onto a set of input values, small local
deviations are inherent to polynomial functional forms
\citep[see e.g.,][]{rm05}. When deriving $\teff$ from colour relations, users
should be mindful of the trade-off between choosing the colour index(es) with
the highest precision versus using as many indices as possible to average down
systematic errors (often at the cost of precision). If one were to use
the mean $\teff$ from all indices, the mean difference with respect to the
spectroscopic measurements in Figure \ref{fig:nb} would be $4\pm 5$~K with a
standard deviation $\sigma=48$~K.

For the GBS we use $\teff, \logg$ and $\feh$ from the
latest version of the catalog \citep{jofre18}. The number of stars with good
photometry varies depending on the filter used, with many of the GBS often
having unreliable or saturated Gaia and/or 2MASS magnitudes. All GBS in our
sample are closer than $\simeq 130$ pc, justifying the adoption of zero
reddening.
Again, we find overall excellent agreement between the $\teff$ we predict from
colours, and those given in the GBS catalog. 

Finally, we assemble a list of interferometric measurements from the recent
literature: \cite{bigot11}, \cite{boya12,boyaM}, \cite{huber12},
\cite{maestro13}, \cite{w13,w18}, \cite{galle18}, \cite{baines18},
\cite{rains20} and \cite{karo20}. For all these
stars, we adopt reddening, $\logg$ and $\feh$ reported in the above papers.
This list encompasses over 200 targets, although most of them are very bright,
hence with unreliable Gaia and/or 2MASS magnitudes, reducing the sample usable
for our comparison to at most thirty-three targets, depending on the colour
index.
For M dwarfs we only retain stars with $(BP-RP)_0 \le 2$ since this is roughly
the reddest colour of dwarfs in GALAH DR3. Note that giant stars go to redder
colours (up to $2.5$, cf.\,Figure \ref{fig:bprp}), although interferometric
$\teff$ of giants are available only for warmer temperatures. For the
comparison in Table
\ref{tab:check} we further require interferometric $\teff$ to be better than 1
percent, which is the target accuracy at which we aim in testing. Allowing for
larger uncertainties results in an increase of scatter in the comparison, with
a trend whereby interferometric $\teff$ are systematically cooler for those
stars with the largest uncertainties. This is indicative that systematic
errors tend to over-resolve angular diameters, hence under-predict effective
temperatures \citep[see discussion in][]{c14}. Also, interferometric targets
with the largest $\teff$ uncertainties are often affected by relatively high
values of reddening, which adds to the error budget. 

\begin{figure*}
\begin{center}
\includegraphics[width=1\textwidth]{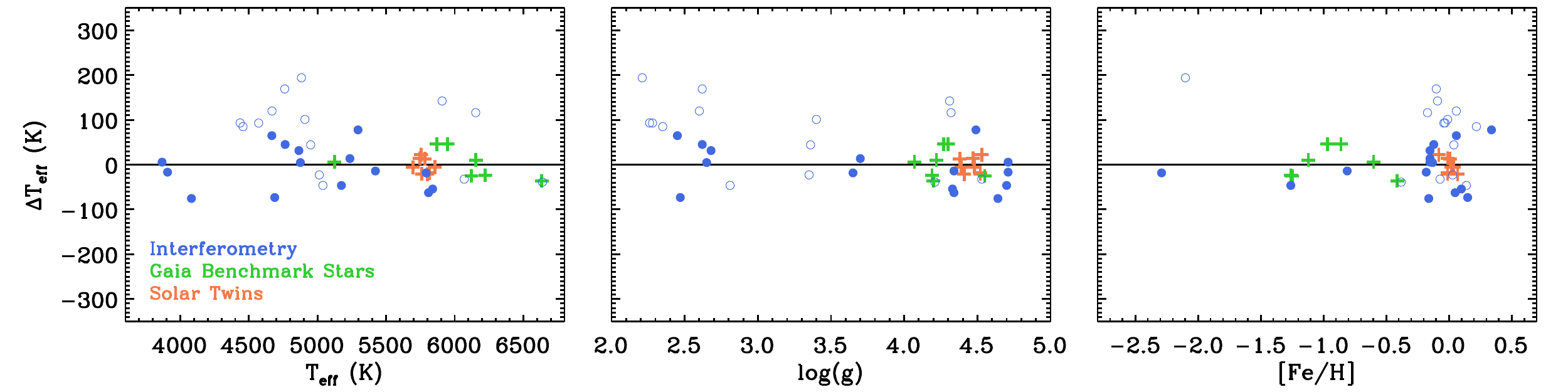}
\caption{Comparison between $\teff$ derived using our $(BP-RP)_0$ relation
  and those available (ours$-$literature) for solar twins (orange), Gaia
  Benchmark Stars (blue) and interferometry (green). Filled and open circles
  indicate interferometric $\teff$ better than 1 and 2 percent,
  respectively.}\label{fig:benc} 
\end{center}
\end{figure*}

Overall, it is clear from Table \ref{tab:check} that our relations are able to
predict $\teff$ values in very good agreement with those reported in the
literature for
various benchmark samples. Depending on the colour index, mean differences are
typically of order few tens of K. Occasional larger differences are still
within the scatter of the relations, or are likely the result of small number
statistics. When we restrict our analysis to the $(BP-RP)_0$ colour index,
which has the largest number of stars available for comparison, the mean
agreement is always within a few K regardless of the sample used (Figure
\ref{fig:benc}).

\begin{figure*}
\begin{center}
\includegraphics[width=1\textwidth]{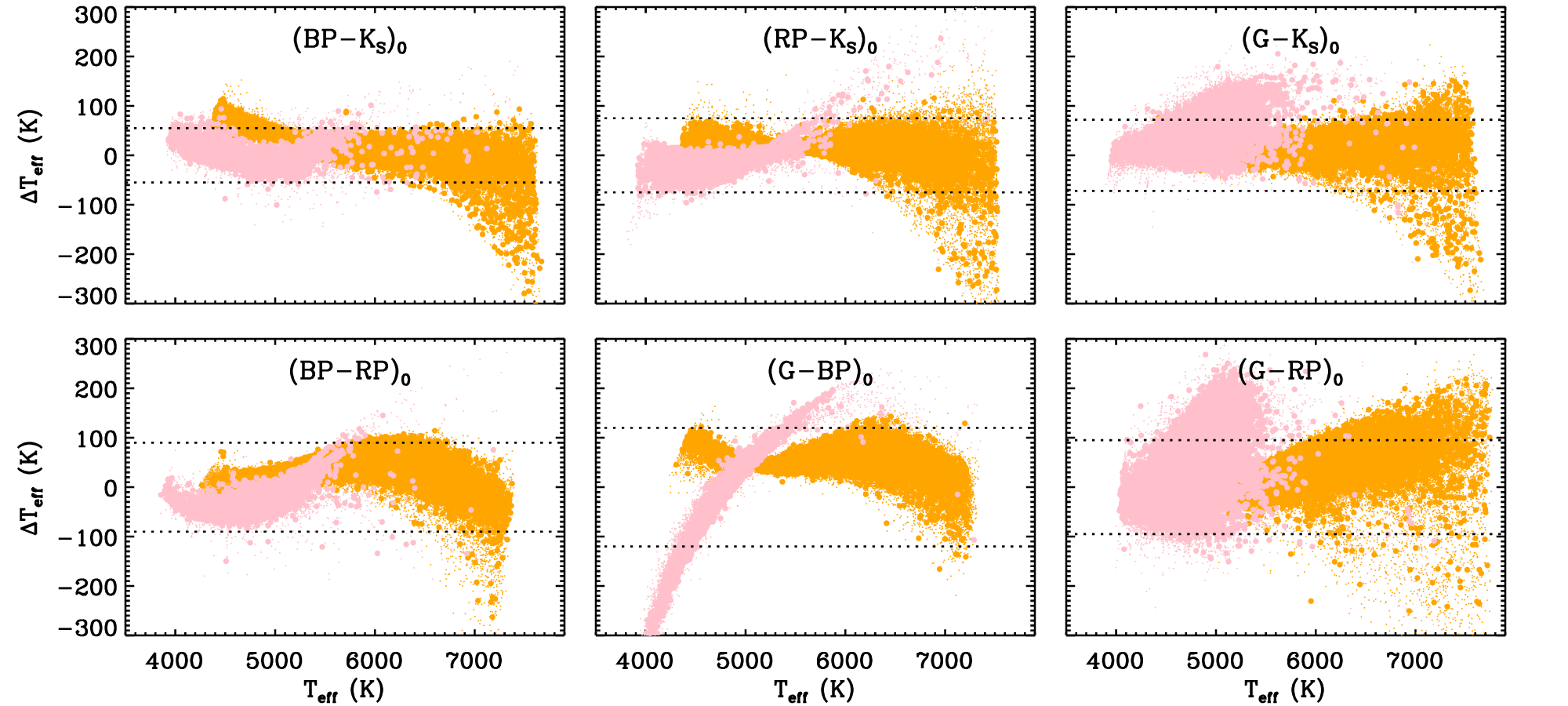}
\caption{Comparison between $\teff$ derived using our relations and those of
  \citet[][MB20]{mb20} in the sense (ours$-$MB20). The
  relations of MB20 do not account for $\logg$, but are provided separately
  for dwarf (orange) and giant (pink) stars. Here, we use the separation
  (dwarfs vs. giants) defined by the dashed line of Figure \ref{fig:test} and
  apply the relations
  within their colour range. The same input $\feh$ and dereddened photometry
  are used for both us and MB20. Filled circles are stars for which
  $\teff$ from our relations are within $10$~K of the IRFM, to ensure
  differences are not stemming from the functional form of our polynomials.
  Dotted lines are the squared root of the squared sum of the typical
  uncertainty quoted for each colour-$\teff$ relation.}\label{fig:mb20} 
\end{center}
\end{figure*}

Finally, we compare our relations against those of \cite{mb20}, which are
the only ones also available for dwarf and giant stars
in the Gaia DR2 system. The colour-$\teff$ relations of \cite{mb20} are built
using several hundred stars with $\teff$ derived from the IRFM work of
\cite{ghb09}. For dwarf stars, the $\teff$ scale of \cite{ghb09}
agrees well with that of
\citet[][which underpins our study]{c10}, with a nearly constant offset of
$30-40$~K (our scale being hotter) due to the
different photometric absolute calibrations adopted. The same offset is
thus expected for \cite{mb20}. This is explored in
Figure \ref{fig:mb20}, which shows the difference between the effective
temperatures obtained using our relations against those of \cite{mb20} for
colour indices in common. The first thing to notice is that the
difference is not a constant offset, but varies as function of $\teff$,
evolutionary status (dwarfs or giants) and colour index. To ensure this
trend does not stem from the functional form of our polynomials, we have
highlighted with filled circles stars for which our colour relations reproduce
input $\teff$ from our IRFM to within 10~K. If one were to take the mean
offset, it would typically be around few tens of K, with a maximum
  of order 50~K for $(G-RP)_0$ and $(G-BP)_0$, our scale being hotter.
Overall, for most stars and colour indices, $\teff$ from our relations agree
with those from \cite{mb20} to within $\sim100$~K, which is the uncertainty
expected
when combining the precision (standard deviation) reported for both
calibrations. Indices with short colour baseline like $(G-RP)_0$ or $(G-BP)_0$
display stronger systematic trends, in particular giants in $(G-BP)_0$.
Larger deviations are also seen around and above 7000~K for dwarf stars,
likely due to the paucity of hot stars available to \cite{mb20} to constrain
well their calibration at high temperatures. Part of the trends might
  also arise from the fact than many of the calibrating giants in \cite{mb20}
  have $G<6$, a regime where Gaia DR2 photometry is affected by uncalibrated
  systematics. For our relations, we have also corrected the standardisation
  of Gaia DR2 $G$
  magnitudes following \cite{maw18}. \cite{mbm21} provide updated relations
using Gaia EDR3 photometry. As discussed in Appendix \ref{appA}, there are
  only minor
  differences between Figure \ref{fig:mb20} and \ref{fig:mbm21} for $(BP-RP)_0$,
  $(BP-K_s)_0$, $(RP-K_s)_0$. This is not surprising, given the overall
  agreement of the colour-$\teff$ relations for Gaia DR2 and EDR3. However,
  indices involving $G$ magnitudes display reduced trends, which in part might 
  arise from the better standardization of $G$ band photometry in EDR3.

From a user point of view, it is important to have realistic estimates of
the precision at which $\teff$ can be estimated from our relations. 
In Table \ref{tab:fit}, we report two values for the standard deviation of our
colour-$\teff$ relations. The first value is the precision of the fits. The
second provides a more realistic assessment of the uncertainties
encountered when applying our relations, and is obtained by randomly
perturbing the input $\feh$ and $\logg$ with a Gaussian distribution of width
$0.2$ and $0.5$ dex, respectively. The effect of a systematic shift of the
GALAH $\logg$ and $\feh$ scale by $\pm0.2$ and $\pm0.1$~dex respectively is
typically also of a few tens of K at most. It should be kept in mind that
uncertainties in the input stellar parameters will propagate differently with
different colours, the effect being strongest for the coolest stars. Users
of our calibrations are encouraged to assess their uncertainties on a
case-by-case basis, by propagating the errors in their input parameters through
Eq.\,\ref{xx}. Further, an extra uncertainty of 20 K should still be added to
account for the zero-point uncertainty of our $\teff$ scale (from
\citealt{c10}, see discussion in Section \ref{sec:irfm}). We provide
  the code {\tt colte}\footnote{\href{https://github.com/casaluca/colte}{\textcolor{blue}{https://github.com/casaluca/colte}}} to derive $\teff$ from our colour relations, taking into account the applicability ranges of Table \ref{tab:range}, and with the option to derive realistic uncertainties through a MonteCarlo for each colour index. Other notable options include the choice of different extinction laws, and Gaia DR2 or EDR3 photometry.

\begin{figure}
\begin{center}
\includegraphics[width=0.5\textwidth]{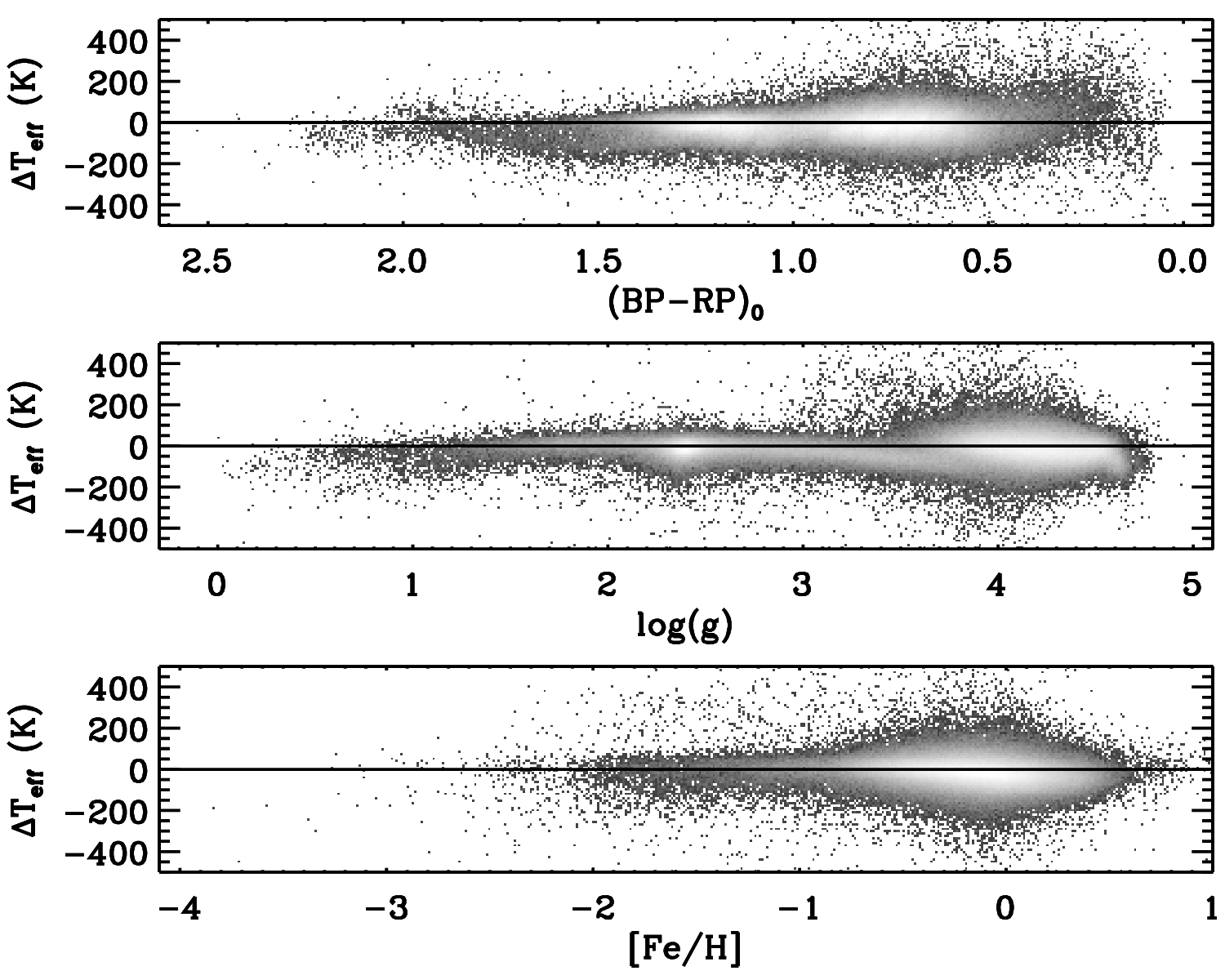}
\caption{$\teff$ residual for the $(BP-RP)_0$ calibration when stars are
  assigned a fixed $\logg=2$ or $4$ based on their classification as giants
  or dwarfs as per Figure \ref{fig:test}. Plots for the other
  colour indices are available as supplementary online material.}\label{fig:bg} 
\end{center}
\end{figure}

Although our calibrations take into account the effect of surface gravity,
there might be instances where the input $\logg$ is not known, besides a
rough ``dwarf'' vs ``giant'' classification. To assess this impact, we classify
stars as dwarfs (giants) if their gravities are higher (lower) than the dashed
line of Figure \ref{fig:test}a. We then adopt a constant $\logg=4$ for dwarfs
and $\logg=2$ for giants. The effect of such an assumption on the derived
$\teff$ is typically small, as can be seen in Figure \ref{fig:bg}.
The largest differences occur for stars in the upper giant branch, where
assuming a constant $\logg=2$ becomes inappropriate for $\logg \lesssim 1-1.5$.
This effect can be quite strong for certain colour indices. In this case, one
might use the
fact that there is a strong correlation between the intrinsic colour and the
surface gravity of stars along the RGB for a better assignment of $\logg$. 

\begin{table}
  \caption{Mean difference and standard deviation between the effective
    temperatures derived from our calibrations, and those from the literature 
    used for validation (ours$-$literature). $N$ is the number of stars
    available in each colour index.}\label{tab:check}
\begin{tabular}{crcrcrc}
\hline
colour & \multicolumn{2}{r}{Solar Twins} & \multicolumn{2}{c}{GBS} & \multicolumn{2}{c}{Interferometry$^{\dag}$}\\
       & $\langle\Delta\teff\rangle$ & N & $\langle\Delta\teff\rangle$ & N & $\langle\Delta\teff\rangle$ & N\\ 
\hline
$(BP-RP)_0$ & $ -3 \pm 17$ & 8 & $  3 \pm  34$ & 7 & $ -8 \pm  49$ & 15 \\
$(G-BP)_0$  & $-30 \pm 16$ & 8 & $-19 \pm  42$ & 5 & $-10 \pm  68$ & 7 \\
$(G-RP)_0$  & $  0 \pm 22$ & 8 & $ 10 \pm  31$ & 5 & $-12 \pm  55$ & 7 \\
$(BP-J)_0$  & $ -6 \pm 23$ & 8 & $ 10 \pm  55$ & 5 & $  9 \pm  48$ & 3 \\
$(BP-H)_0$  & $ 10 \pm 13$ & 8 & $ 49 \pm  48$ & 5 & $ 95 \pm   8$ & 2 \\
$(BP-K_s)_0$  & $-14 \pm 21$ & 8 & $-16 \pm  32$ & 6 & $-30 \pm  60$ & 6 \\
$(RP-J)_0$  & $-13 \pm 69$ & 8 & $  2 \pm 108$ & 5 & $-42 \pm 112$ & 3 \\
$(RP-H)_0$  & $ -2 \pm 36$ & 8 & $ 25 \pm  74$ & 5 & $ 68 \pm  14$ & 2 \\
$(RP-K_s)_0$  & $-28 \pm 37$ & 8 & $-52 \pm  40$ & 6 & $-51 \pm  59$ & 6 \\
$(G-J)_0$   & $ -2 \pm 36$ & 8 & $ -4 \pm  72$ & 5 & $-16 \pm  49$ & 3 \\
$(G-H)_0$   & $  5 \pm 20$ & 8 & $ 18 \pm  59$ & 5 & $ 60 \pm  50$ & 2 \\
$(G-K_s)_0$   & $-27 \pm 26$ & 8 & $-52 \pm  40$ & 5 & $-54 \pm  74$ & 5 \\
\hline
\end{tabular}
\begin{minipage}{1.4\textwidth}
$^{\dag}$Only interferometric $\teff$ better than 1 percent are used. 
\end{minipage}
\end{table}

\section{Conclusions}\label{sec:finale}

In this paper we have implemented the Gaia DR2 and EDR3 photometric system in
the IRFM and applied to over 360,000 stars with good spectroscopic and
photometric flags to derive $\teff$ for stars across
different evolutionary phases. In
the literature, colour-$\teff$ relations for late type-stars are typically
given separately for dwarfs and giants. The advent of Gaia parallaxes allows
us to use robust surface gravities together with $\feh$ from the GALAH DR3
survey to provide colour-$\teff$ relations that take into account the effect
of these two parameters. Our calibrations are built and tested using the largest
high-resolution stellar spectroscopic survey to date and cover a wide range
of stellar colours and parameters: $0 \lesssim \logg \lesssim 4.8$ and
$-3 \lesssim \feh \lesssim 0.6$. When using our relations, users should
refer to Figures \ref{fig:bprp} and \ref{fig:test} to have a sense for
the parameter space covered, and for the performances of different
colour indices. Users should always be mindful of the 
  trade-off between choosing the colour index(es) with the highest precision
  versus using as many indices as possible to average down systematic errors,
  often at the cost of precision. $(BP-K_s)_0$ and $(G-K_s)_0$ are the
indices which
are best calibrated against $\teff$ across the parameter space, whereas
indices leveranging on $RP$ are the least performing ones. In particular,
$(RP-J)_0$ has a very short colour baseline and the largest scatter, and
other colour indices should be used instead, if possible. Moving
to indices built only with Gaia filters, $(BP-RP)_0$ is the best choice,
although $(G-BP)_0$ and $(G-RP)_0$ are also informative. For solar twins,
all three indices return $\teff$ with remarkably small scatter with respect to
the highly precise ones derived from differential spectroscopic analyses.
Robust solar colours have also been derived (Appendix \ref{appC}). For most
colour indices, our calibrations have a typical 1 sigma uncertainty
of $40-80$~K for the colour intervals of Table \ref{tab:range}, which cover
the region between 4000~K and 8000~K. For
$4000\,\rm{K}\lesssim \teff \lesssim 6700$\,K our calibrations are also
validated against solar twins, Gaia Benchmark Stars and interferometry.

\section*{Data availability}
The data underlying this article were accessed from the GALAH survey DR3 which
can be queried using TAP at \href{https://datacentral.org.au/vo/tap}{https://datacentral.org.au/vo/tap}. The derived data generated in this research will be shared on reasonable request to the corresponding author.

\setcounter{table}{0}
\begin{landscape}
\thispagestyle{empty}
  \begin{table}
\caption{Coefficients of the $\teff$ calibration of Eq.\,\ref{xx} suitable for Gaia DR2 photometry. See Appendix \ref{appA} for Gaia EDR3 photometry.}\label{tab:fit}
\begin{tabular}{ccccccccccccccccc}
\hline
colour & $a_0$ & $a_1$ & $a_2$ & $a_3$ & $a_4$ & $a_5$ & $a_6$ & $a_7$ & $a_8$ & $a_9$ & $a_{10}$ & $a_{11}$ & $a_{12}$ & $a_{13}$ & $a_{14}$ & $\sigma(\teff)$\,(K) \\
\hline
$(BP-RP)_0$ & $7928$  & $-3663.1140$   &  $803.3017$  &    $-9.3727$ &     $-$      &  $325.1324$ &   $-500.1160$ &  $279.4832$ &  $-53.5062$ &   $-$      &   $-2.4205$ &  $-128.0354$ & $49.4933$ &  $5.9146$ &   $41.3650$ &  $54 - 66$ \\
$(G-BP)_0$  & $7555$  & $ 5803.7715$   &     $-$      & $-2441.7124$ &   $437.7314$ &  $455.0997$ &   $2243.1333$ & $3669.4924$ & $1872.7035$ &   $-$      &   $19.1085$ &    $75.2198$ &  $-$      &  $-$      &  $-83.9777$ &  $75-93$ \\
$(G-RP)_0$  & $7971$  & $-5737.5049$   &     $-$      &  $1619.9946$ &  $-203.8234$ &  $255.7408$ &   $-492.8268$ &  $160.1957$ &  $103.1114$ &   $-$      &  $-64.3289$ &    $34.3339$ &  $-$      &  $-$      &   $54.7224$ &  $56-64$ \\
$(BP-J)_0$  & $8218$  & $-2526.8430$   &  $458.1827$  &   $-28.4540$ &     $-$      &  $234.0113$ &   $-205.3084$ &   $63.4781$ &   $-7.2083$ &   $-$      &  $-85.7048$ &   $-50.1557$ & $32.3428$ & $-2.3553$ &   $20.0671$ &  $44-49$ \\
$(BP-H)_0$  & $8462$  & $-2570.3684$   &  $537.5968$  &   $-44.3644$ &     $-$      &  $189.1198$ &   $-106.7584$ &   $31.1720$ &   $-4.9137$ &   $-$      &   $-9.2587$ &  $-189.8600$ & $75.8619$ & $-6.8592$ &   $16.7226$ &  $33-42$ \\
$(BP-K_s)_0$  & $8404$  & $-2265.1355$   &  $403.4693$  &   $-27.9056$ &     $-$      &  $193.5820$ &   $-145.3724$ &   $47.7998$ &   $-6.4572$ &   $-$      &  $-34.5438$ &  $-130.2559$ & $52.6470$ & $-4.4777$ &   $15.8249$ &  $24-32$ \\ 
$(RP-J)_0$  & $9074$  & $-7670.6606$   & $3164.0525$  &     $-$      &  $-126.1476$ &    $-$      &     $-7.3816$ &  $-12.5168$ &    $-$      &  $-2.0452$ &    $-$      &    $76.1144$ &  $-$      &  $-$      &  $-45.8056$ &  $90-95$ \\
$(RP-H)_0$  & $8924$  & $-4779.3394$   & $1319.8989$  &     $-$      &   $-16.6676$ &    $-$      &    $-23.6583$ &   $22.4243$ &    $-$      &  $-4.3066$ &    $-$      &    $35.0102$ &  $-$      &  $-$      &  $-28.7228$ &  $52-62$ \\
$(RP-K_s)_0$  & $8940$  & $-4450.6138$   & $1138.6816$  &     $-$      &   $-10.5749$ &    $-$      &    $-42.3037$ &   $33.3365$ &    $-$      &  $-3.2535$ &    $-$      &    $41.0402$ &  $-$      &  $-$      &  $-21.9922$ &  $43-48$ \\
$(G-J)_0$   & $8370$  & $-3559.7710$   &  $895.8869$  &   $-86.7011$ &     $-$      &  $180.7568$ &   $-164.9264$ &   $24.4263$ &    $4.2318$ &   $-$      & $-127.9640$ &    $72.1449$ &  $-$      &  $-$      &   $13.7683$ &  $54-57$ \\  
$(G-H)_0$   & $8186$  & $-2536.7671$   &  $503.2762$  &   $-42.7871$ &     $-$      &  $230.4871$ &   $-254.5291$ &  $104.6258$ &  $-17.4859$ &   $-$      & $-122.0732$ &    $45.0572$ &  $-$      &  $-$      &    $6.9992$ &  $37-41$ \\
$(G-K_s)_0$   & $8103$  & $-1857.7194$   &     $-$      &    $73.1834$ &    $-1.7576$ &  $236.0335$ &   $-345.9070$ &  $170.4915$ &  $-28.8549$ &   $-$      & $-131.4548$ &    $49.6232$ &  $-$      &  $-$      &   $10.0777$ &  $27-32$ \\
\hline
\end{tabular}
\begin{minipage}{1.4\textwidth}
  $G$ magnitudes have been corrected following \cite{maw18}: $G+0.0271\,(6-G)$
  for $G\le6$, $G-0.0032\,(G-6)$ for $6<G<16$ and $G-0.032$ for $G\ge16$. 
  These corrections should be applied before using our relations with $G$
  magnitudes: the effect on indices with short colour baseline such as
  $(G-BP)_0$ and $(G-RP)_0$ is noticeable, and up to $100-200$~K for hot stars
  in particular. See Table \ref{tab:exte} for extinction coefficients suitable
  for Gaia DR2 and 2MASS. 
  Users should also be wary of applying colour-$\teff$ relations to stars with
  $G<6$
  and $BP$ and $RP$ brighter than $\sim 5$ due to the saturation of bright
  magnitudes in Gaia. For the standard deviation of the calibration
  $\sigma(\teff)$, we provide
two estimates, both obtained using all available $\sim 360,000$ stars, instead
of the $\sim 50,000$ used to derive fits. The first one is the precision of
the fits, whereas for the second one input $\feh$ and $\logg$ are perturbed
with  a Gaussian random noise of $0.2$ and $0.5$ dex, respectively. Note that an
extra uncertainty of about 20 K on the zero-point of our effective temperature
scale should still be added. 
\end{minipage}
\end{table}
\end{landscape} 

\section*{Acknowledgments}

We thank the referee for their valuable comments and suggestions. 
LC is the recipient of an ARC Future Fellowship (project number FT160100402).
ADR acknowledges support from the Australian Government Research Training
Program, and the Research School of Astronomy \& Astrophysics top up
scholarship. SLM acknowledges support from the UNSW Scientia Fellowship program.
SLM, JS and DZ acknowledge support from the Australian Research Council through
Discovery Project grant DP180101791. YST is grateful to be supported by the
NASA Hubble Fellowship grant HST-HF2-51425.001 awarded by the Space Telescope
Science Institute. JK and TZ acknowledge funding from the Slovenian Research
Agency (grant P1-0188).
Parts of this research were conducted by the Australian Research Council Centre
of Excellence for All Sky Astrophysics in 3 Dimensions (ASTRO 3D), through
project number CE170100013. This work has made use of data from the European
Space Agency (ESA) mission {\it Gaia} (\url{https://www.cosmos.esa.int/gaia}),
processed by the {\it Gaia} Data Processing and Analysis Consortium (DPAC,
\url{https://www.cosmos.esa.int/web/gaia/dpac/consortium}). Funding
for the DPAC has been provided by national institutions, in particular
the institutions participating in the {\it Gaia} Multilateral Agreement.
\vspace{-0.7cm}
\bibliographystyle{mn2e}
\bibliography{refs}

\appendix

\section[]{Colour-$\teff$ relations using Gaia EDR3 photometry}\label{appA}

The IRFM and colour-$\teff$ relations described in the paper are based on Gaia
DR2 photometry. Here we discuss the implementation of Gaia EDR3 photometry into
the IRFM and provide colour-$\teff$ relations for this system.
\begin{figure*}
\begin{center}
\includegraphics[width=1\textwidth]{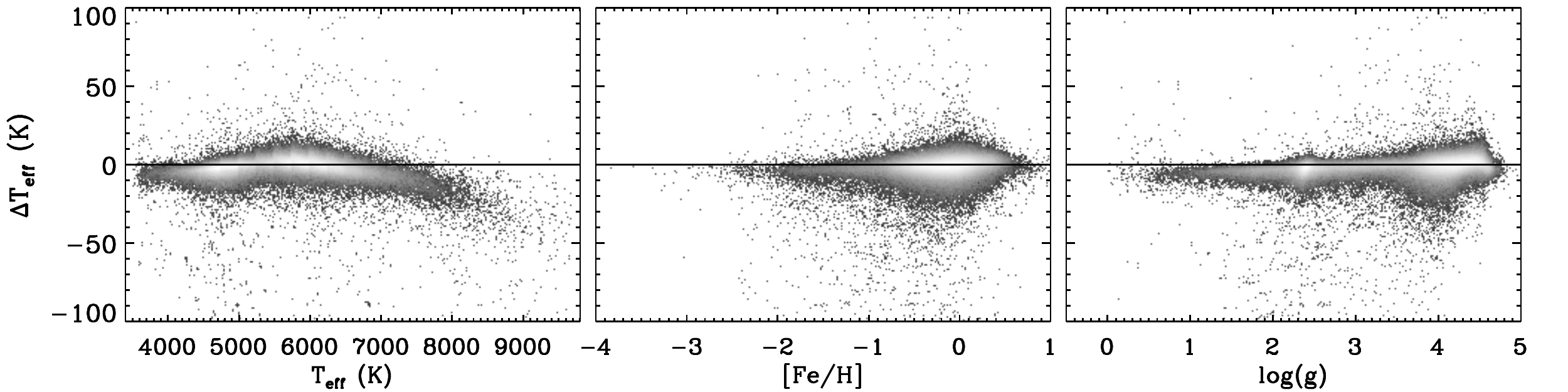}
\caption{Log-density plot of the difference in effective temperatures
    derived by the IRFM when implementing Gaia EDR3 photometry
  instead of DR2 in the optical (EDR3 minus DR2). Approximately 355,000
  stars
  with good GALAH spectroscopic, and photometric flags in both EDR3 and DR2
  are shown here. For 96 percent of the stars the difference is always within
  $\pm10$~K.}\label{fig:app1} 
\end{center}
\end{figure*}

Gaia EDR3 photometry defines an independent photometric system from Gaia DR2,
with significant improvements in the processing of the data and photometric
calibration \cite[see][for an in depth discussion]{Riello20}. These improvements
affect not only the published EDR3 magnitudes (and fluxes), but also the
filter transmission curves and zero-points defining the system. Here, we
implement EDR3 passbands and zero-points, along with EDR3 $BP$ and $RP$
photometry into the IRFM. As described in Section \ref{sec:irfm}, 2MASS
  $JHK_s$ are used in the infrared.
Also in this instance we do not use the redundant
information from EDR3 $G$ magnitudes in the IRFM, although we do provide
calibrations involving this band. $BP$, $RP$ and $G$ magnitudes for bright
sources have been corrected for saturation effects following \cite{Riello20}.
$G$ magnitude correction for bright blue sources is not applied since none of
our target is bluer than $BP-RP\sim0$, but we correct $G$ magnitudes for sources with 2 or 6-parameter astrometric solutions\footnote{\href{https://github.com/agabrown/gaiaedr3-6p-gband-correction}{\textcolor{blue}{https://github.com/agabrown/gaiaedr3-6p-gband-correction}}}. 
\begin{figure*}
\begin{center}
\includegraphics[width=1\textwidth]{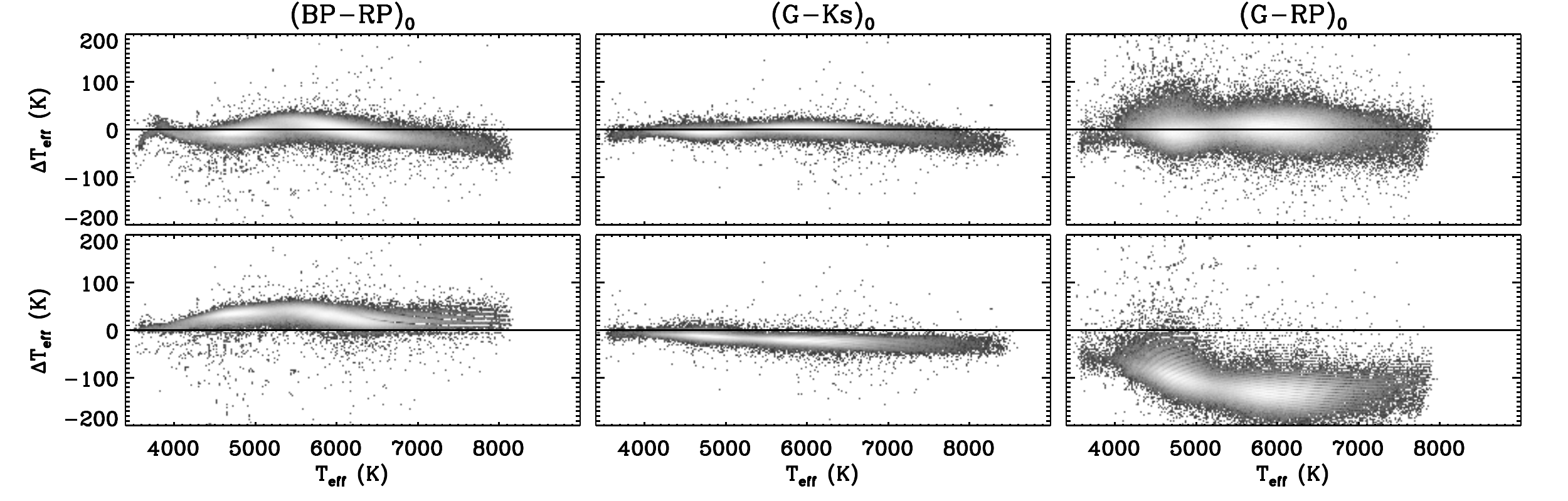}
\caption{Top panels: log-density plots of the
  effective temperature difference between the Gaia EDR3 and DR2 calibration
  when photometry from the corresponding release is used. Bottom panels: 
  effective temperature difference between using both EDR3 and DR2 photometry
  into the EDR3 calibration. The offset in $(G-RP)_0$ largely originates from
  the correction applied to DR2 $G$ magnitudes (see discussion in the text).
  In all instances, calibrations have been applied within the validity ranges
  of Table \ref{tab:range}. }\label{fig:app2} 
\end{center}
\end{figure*}

As in Section \ref{sec:irfm}, we derive $\teff$ for all stars in \cite{c10}
with a counterpart in EDR3 (now 410 targets), obtaining a mean and median
$\Delta\teff=17\pm2$~K ($\sigma=41$~K). The mean $\teff$ difference of
implementing Gaia EDR3 instead of DR2 photometry is a mere 5K with a slight
trend as function of $\teff$. The latter is more clearly visible when comparing
effective temperatures obtained from the IRFM for the entire GALAH sample
(Figure \ref{fig:app1}). For 96 (99) percent of stars the difference is always
within $\pm10$~K ($\pm20$~K), well within the zero-point uncertainty of our
scale, and no noticeable trends with surface gravity and metallicity. Above
7500~K
however there is the tendency for EDR3 to return effective temperatures which
are systematically cooler by some tens of K.

Table \ref{tab:app} provides colour-$\teff$ coefficients derived in a similar
fashion to Table \ref{tab:fit}, but using instead EDR3 photometry. We select
good photometry by requesting {\tt phot\_proc\_mode=0} and $BP$ and $RP$
corrected excess factor\footnote{\href{https://github.com/agabrown/gaiaedr3-flux-excess-correction}{\textcolor{blue}{https://github.com/agabrown/gaiaedr3-flux-excess-correction}}} $-0.08<C^{\star}<0.2$ \citep{Riello20}. This last 
requirement is similar to
$0.001+0.039\,(BP-RP)<\log_{10}({\tt phot\_bp\_rp\_excess\_factor})
<0.12+0.039\,(BP-RP)$ used by \cite{anti} to select good photometry.
Note that extinction coefficients for Gaia EDR3 filters are also updated from
Figure \ref{fig:exco}, and provided in Table \ref{tab:exte}.

It is important to note that although the calibration for Gaia DR2 and EDR3
are overall similar, photometry from one system should never be
used with the calibration of the other. The danger of doing this is shown for
a few
selected colour combinations in Figure \ref{fig:app2}. On the top panels, when
photometry from a data release is used with its colour-$\teff$ relation, the
agreement between effective temperatures is consistent to
what expected from Figure \ref{fig:app1} (as calibrations in different indices
have their own intrinsic scatter). However, if -say- photometry
from DR2 is used onto the calibration for EDR3 (equivalent of plotting the
difference of the relations at same colour), systematic offsets will appear.
This is particularly relevant for indices involving $G$ magnitudes, which for
Gaia DR2 have been corrected following \cite{maw18}.  Although this correction
is magnitude
dependent, over the range of our stars it amounts to few hundredths of a
magnitude. This difference does not significantly impact $\teff$ in colours
with long baseline, see e.g., $(G-K_s)_0$ in the bottom mid panel of Figure
\ref{fig:app2}. However, for indices like $(G-BP)_0$ or $(G-RP)_0$, effective
temperatures can be off by as much as $100-200$~K (bottom right panel of
Figure \ref{fig:app2}).

\begin{figure*}
\begin{center}
\includegraphics[width=1\textwidth]{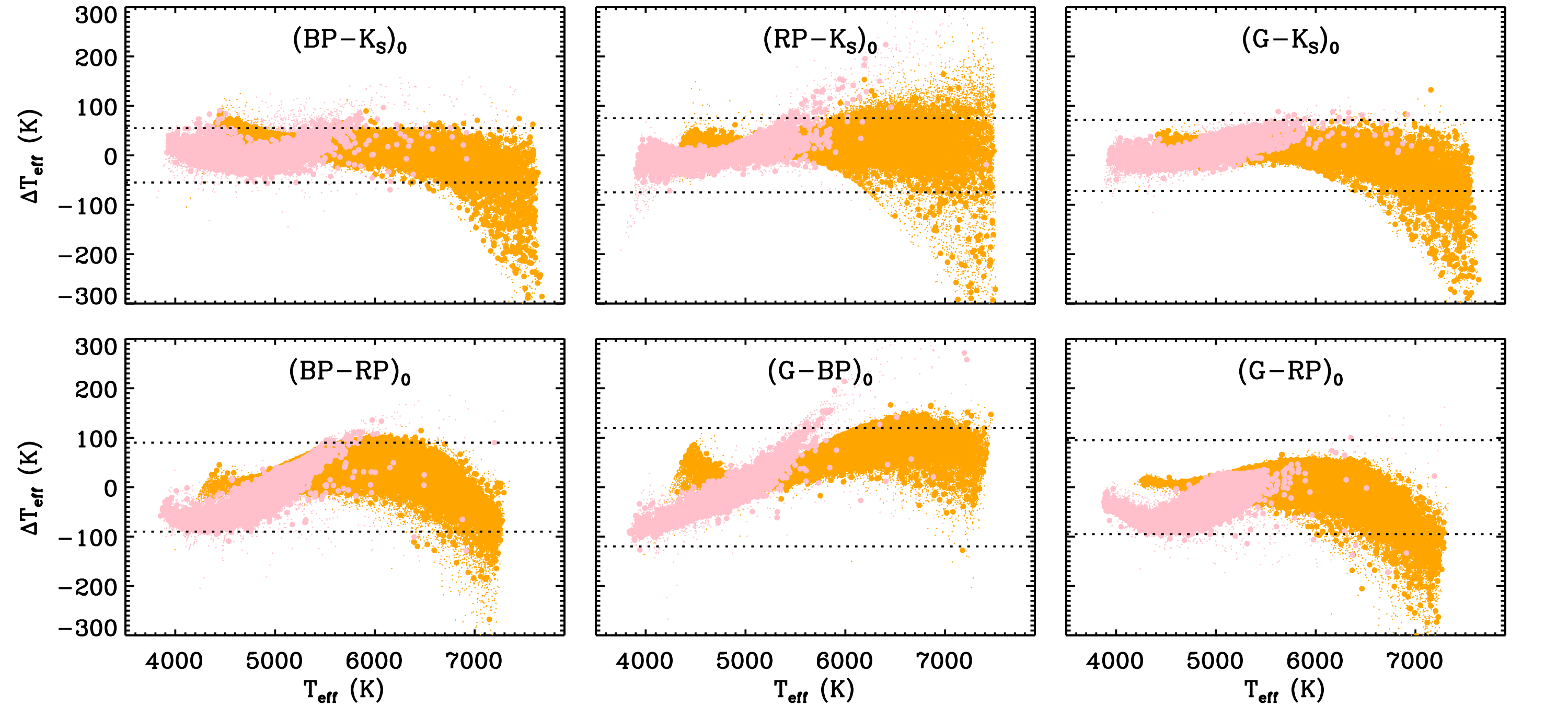}
\caption{Same as Figure \ref{fig:mb20}, but comparing our Gaia EDR3 relations
  against those of \citet{mbm21}.}\label{fig:mbm21} 
\end{center}
\end{figure*}

Although $G$ magnitudes for sources with 2 or 6-parameter astrometric
solutions still need minor corrections in EDR3, zero-point shifts to improve
standardisation are not necessary anymore \citep[cf.][for Gaia DR2]{cv18,maw18}.
Similarly, many of the bright objects used by \cite{mbm21} to define their
relations have much improved $G$ band photometry in EDR3. This
likely explains the reduced trends when comparing our EDR3 calibrations
against those of \cite{mbm21} for indices involving $G$ band (see Figure
\ref{fig:mbm21} and discussion in Section \ref{sec:test}).

\begin{landscape}
\thispagestyle{empty}
  \begin{table}
\caption{Coefficients of the $\teff$ calibration of Eq.\,\ref{xx} suitable for Gaia EDR3 photometry.}\label{tab:app}
\begin{tabular}{ccccccccccccccccc}
\hline
colour & $a_0$ & $a_1$ & $a_2$ & $a_3$ & $a_4$ & $a_5$ & $a_6$ & $a_7$ & $a_8$ & $a_9$ & $a_{10}$ & $a_{11}$ & $a_{12}$ & $a_{13}$ & $a_{14}$ & $\sigma(\teff)$\,(K) \\
\hline
$(BP-RP)_0$  & $7981$  & $-4138.3457$ &   $1264.9366$ &  $-130.4388$  &     $-$     &  $285.8393$ &  $-324.2196$ &  $106.8511$ &   $-4.9825$ &    $-$     &   $4.5138$  &  $-203.7774$ &  $126.6981$ & $-14.7442$ &   $40.7376$ & $55-64$ \\
$(G-BP)_0$   & $7346$  & $ 5810.6636$ &       $-$     & $-2880.3823$  &  $669.3810$ &  $415.3961$ &  $2084.4883$ & $3509.2200$ & $1849.0223$ &    $-$     &  $-49.0748$ &     $6.8032$ &     $-$     &    $-$     & $-100.3419$ & $76-88$ \\
$(G-RP)_0$   & $8027$  & $-5796.4277$ &       $-$     &  $1747.7036$  & $-308.7685$ &  $248.1828$ &  $-323.9569$ & $-120.2658$ &  $225.9584$ &    $-$     &  $-35.8856$ &   $-16.5715$ &     $-$     &    $-$     &   $48.5619$ & $53-61$ \\
$(BP-J)_0$   & $8172$  & $-2508.6436$ &    $442.6771$ &  $ -25.3120$  &     $-$     &  $251.5862$ &  $-240.7094$ &   $86.0579$ &  $-11.2705$ &    $-$     &  $-45.9166$ &  $-137.4645$ &   $75.3191$ &  $-8.7175$ &   $21.5739$ & $44-49$ \\
$(BP-H)_0$   & $8159$  & $-2146.1221$ &    $368.1630$ &  $ -24.4624$  &     $-$     &  $231.8680$ &  $-170.8788$ &   $52.9164$ &   $-6.8455$ &    $-$     &  $-45.5554$ &  $-142.9127$ &   $55.2465$ &  $-4.1694$ &   $17.6593$ & $32-40$ \\
$(BP-K_s)_0$ & $8266$  & $-2124.5574$ &    $355.5051$ &  $ -23.1719$  &     $-$     &  $209.9927$ &  $-161.4505$ &   $50.5904$ &   $-6.3337$ &    $-$     &  $-27.2653$ &  $-160.3595$ &   $67.9016$ &  $-6.5232$ &   $16.5137$ & $24-33$ \\
$(RP-J)_0$   & $9047$  & $-7392.3789$ &   $2841.5464$ &     $-$       &  $-85.7060$ &     $-$     &   $-88.8397$ &   $80.2959$ &     $-$     & $-15.3872$ &    $-$      &    $54.6816$ &     $-$     &    $-$     &  $-32.9499$ & $91-93$ \\
$(RP-H)_0$   & $8871$  & $-4702.5469$ &   $1282.3384$ &     $-$       &  $-15.8164$ &     $-$     &   $-30.1373$ &   $27.9228$ &     $-$     &  $-4.8012$ &    $-$      &    $25.1870$ &     $-$     &    $-$     &  $-22.3020$ & $52-59$ \\
$(RP-K_s)_0$ & $8911$  & $-4305.9927$ &   $1051.8759$ &     $-$       &   $-8.6045$ &     $-$     &   $-76.7984$ &   $55.5861$ &     $-$     &  $-3.9681$ &    $-$      &    $35.4718$ &     $-$     &    $-$     &  $-16.4448$ & $43-46$ \\
$(G-J)_0$    & $8142$  & $-3003.2988$ &    $499.1325$ &    $-4.8473$  &     $-$     &  $244.5030$ &  $-303.1783$ &  $125.8628$ &  $-18.2917$ &    $-$     & $-125.8444$ &    $59.5183$ &     $-$     &    $-$     &   $16.8172$ & $53-56$ \\
$(G-H)_0$    & $8134$  & $-2573.4998$ &    $554.7657$ &   $-54.0710$  &     $-$     &  $229.2455$ &  $-206.8658$ &   $68.6489$ &  $-10.5528$ &    $-$     & $-124.5804$ &    $41.9630$ &     $-$     &    $-$     &    $7.9258$ & $36-41$ \\
$(G-K_s)_0$  & $8032$  & $-1815.3523$ &       $-$     &    $70.7201$  &   $-1.7309$ &  $252.9647$ &  $-342.0817$ &  $161.3031$ &  $-26.7714$ &    $-$     & $-120.1133$ &    $42.6723$ &     $-$     &    $-$     &   $10.0433$ & $27-31$ \\
\hline
\end{tabular}
\begin{minipage}{1.4\textwidth}
  Refer to Table \ref{tab:fit} for a description of the columns. The same colour
  limits given in Table \ref{tab:range} apply here. Before using these
  relations, $G, BP$ and $RP$ magnitudes for bright sources needs to be
  corrected for saturation. For sources with 2 or 6-parameter astrometric
  solutions
  $G$ magnitudes must also be corrected \citep{Riello20}. See Table
  \ref{tab:exte} for extinction coefficients suitable for Gaia EDR3 and 2MASS.
\end{minipage}
\end{table}
\end{landscape}

\section[]{The dependence of colour-$\teff$ relations on the adopted extinction law}\label{appB}

The relations of Table \ref{tab:fit} and \ref{tab:app} have been derived
adopting the \cite{ccm89}/\cite{od94} extinction law (hereafer COD) for
consistency with our
earlier work on the IRFM \citep{c10}. Here, we investigate the effect of
using a different extinction law, namely that of \cite{f99}, renormalized
as per \citet[][hereafter referred to as FSF]{sf11}. Changing law affects
the amount of extinction inferred in each photometric band for a given input
$E(B-V)$. In other words, different extinction coefficients will be derived.
This is due to the fact that extinction laws have different
normalizations and shapes. Because of the normalization, extinction
coefficients will be higher or lower by a similar percent. Because
of the shape, certain photometric bands will be affected more than others in
relative terms. Changes in normalization and shape of extinction laws can also
be due to variations in $R_V$ (i.e. the ratio of total to selective
extinction in $V$ band, used to build a one-parameter family of curves).
In this work, however, we adopt the ``standard'' $R_V=3.1$ which
applies to the diffuse interstellar medium for most line of sights in the
Galaxy. 

\begin{figure*}
\begin{center}
\includegraphics[width=1\textwidth]{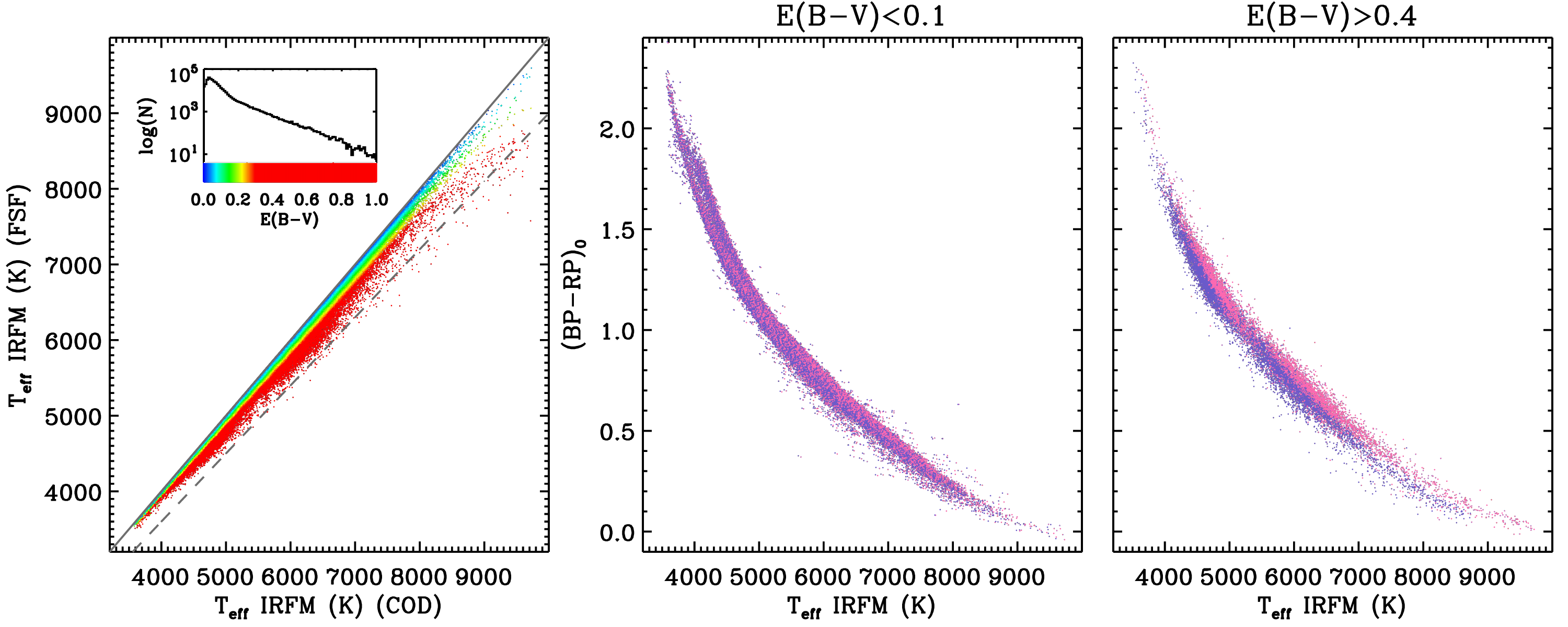}
\caption{Left panel: comparison between $\teff$ obtained implementing the COD
  or the FSF extinction law in the IRFM. Stars are colour-coded by
  their $E(B-V)$ with the distribution shown in the inset. Continuous grey line
  is the one-to-one relation, whereas the dashed line marks a 10 percent
  decrease in $\teff$. Central panel: colour-$\teff$ derived using
  the COD (pink) and the FSF (purple) extinction laws when reddening is below
  $0.1$. Right panel: same as central panel, but when reddening is above
  $0.4$. }\label{fig:app_comp} 
\end{center}
\end{figure*}

Depending on the extinction law, different unreddened colours will be
obtained for the same input reddening, thus affecting photometric effective
temperatures. The extinction coefficients derived with FSF are roughly 15 to
25 percent lower than with COD, implying that $\teff$ of stars affected by
reddening will be cooler assuming the former extinction law (Table
\ref{tab:exte}). This is shown in the left
panel of Figure \ref{fig:app_comp}, which compares $\teff$ derived using the
COD or the FSF law into the IRFM. For the highest reddening
values in our sample the difference in temperature
can reach up to $\sim$10 percent, which corresponds to several hundreads of K
for hot stars. Fortunately, the effect on the colour-$\teff$ relations is much
smaller. For low reddening values (central panel of Figure \ref{fig:app_comp}),
bluer or redder stellar colours map into hotter or cooler effective
temperature, roughly
moving on the same colour-$\teff$ relation, regardless of the underlying
extinction law. Thus, even if our relations have been derived using the COD law,
a change of extinction coefficients suffices to derive effective temperatures
under different extinction curves. This has been verified by using the
coefficients in Table \ref{tab:exte} with the calibrations of Table
\ref{tab:fit} and \ref{tab:app}: within the precision allowed by our
colour-$\teff$ relations,  we are able to recover $\teff$ when the COD or FSD
law is implemented in the IRFM directly.

\begin{table*}
  \caption{Colour dependent extinction coefficients
    $R_\zeta=b_0+b_1(BP-RP)_0+b_2(BP-RP)_0^2+b_3(BP-RP)_0^3$ for Gaia and 2MASS
    photometry assuming different laws. To estimate intrinsic colours needed
    for the fits, one can iterate starting with the assumption
    $(BP-RP)_0 \simeq (BP-RP)-E(B-V)$.}\label{tab:exte}
\begin{tabular}{ccccc|cccc|cccc|cccc}
\hline
       & \multicolumn{8}{c}{COD extinction law} &  \multicolumn{8}{c}{FSF extinction law} \\ 
\hline
       & \multicolumn{4}{c}{Gaia DR2} & \multicolumn{4}{c}{Gaia EDR3} & \multicolumn{4}{c}{Gaia DR2} & \multicolumn{4}{c}{Gaia EDR3} \\ 
\hline
        & $b_0$   &  $b_1$   &  $b_2$  & $b_3$ &  $b_0$  &  $b_1$   &  $b_2$   & $b_3$ & $b_0$ & $b_1$ & $b_2$ & $b_3$ & $b_0$ & $b_1$ & $b_2$ & $b_3$ \\
\hline
$R_{G}$  & $3.068$ & $-0.504$ &  $0.053$ &    $-$   & $3.071$ & $-0.511$ &  $0.058$ &    $-$   & $2.608$ & $-0.468$ &  $0.048$ &   $-$   & $2.609$ & $-0.475$ &  $0.053$ &   $-$   \\ 
$R_{BP}$ & $3.533$ & $-0.114$ & $-0.219$ &  $0.070$ & $3.526$ & $-0.168$ & $-0.170$ &  $0.060$ & $3.007$ & $-0.099$ & $-0.212$ & $0.069$ & $2.998$ & $-0.140$ & $-0.175$ & $0.062$ \\ 
$R_{RP}$ & $2.078$ & $-0.073$ &   $-$    &    $-$   & $2.062$ & $-0.072$ &   $-$    &    $-$   & $1.702$ & $-0.060$ &    $-$   &   $-$   & $1.689$ & $-0.059$ &   $-$    &   $-$   \\ 
$R_{J}$  & $0.899$ &   $-$    &   $-$    &    $-$   & $0.899$ &   $-$    &   $-$    &    $-$   & $0.719$ &   $-$    &    $-$   &   $-$   & $0.719$ &   $-$    &   $-$    &   $-$   \\
$R_{H}$  & $0.567$ &   $-$    &   $-$    &    $-$   & $0.567$ &   $-$    &   $-$    &    $-$   & $0.455$ &   $-$    &    $-$   &   $-$   & $0.455$ &   $-$    &   $-$    &   $-$   \\
$R_{K_{s}}$ & $0.366$ &   $-$    &   $-$    &    $-$   & $0.366$ &   $-$    &   $-$    &    $-$   & $0.306$ &   $-$    &    $-$   &   $-$   & $0.306$ &   $-$    &   $-$    &   $-$   \\
\hline
\end{tabular}
\begin{minipage}{1\textwidth}
See discussion in Appendix \ref{appB} for the definition of COD and FSF extinction laws. 
\end{minipage}
\end{table*}

\section[]{Solar colours}\label{appC}

By fixing the solar surface gravity, metallicity and effective temperature,
Eq.~\ref{xx} can be solved to derive the colours of the Sun. Here we adopt
$\log(g)_{\odot}=4.44$ and $T_{\rm{eff},\odot}=5777$~K, where the latter value is
kept for consistency with our previous sets of solar colours \citep{c10,c12}
We verified however that if we were to adopt the effective temperature
recommended by the IAU 2015 Resolution B3 \citep[5772~K,][]{prsa} the derived
colours would change at most by $0.004$~mag, which is considerably less than
our uncertainties (where a lower $T_{\rm{eff},\odot}$ implies redder solar
colours).

In Table \ref{tab:sun} we report the colours derived from Table
\ref{tab:fit} and \ref{tab:app} for the Gaia DR2 and EDR3 system,
respectively. The precision $\sigma(\teff)$ quoted for our colour-$\teff$
relations is used to perturb $T_{\rm{eff},\odot}$, and to derive uncertainties
for the colours of the Sun. The 20~K uncertainty on the zero-point of our
effective temperature scale is not included, and it would typically imply a
systematic shift to our colours of order 0.01 mag, depending on the index. 

For comparison we also derive solar colours using four high fidelity, flux
calibrated spectra (from  \citealt{rieke08}, the CALSPEC solar reference
spectrum sun\_reference\_stis\_002, and the solar irradiance spectra of
\citealt{thui04} and \citealt{mef18}).
The zero-points and transmission curves used to compute colours from these
spectra are the same we have adopted in the IRFM for the Gaia DR2, EDR3 and
2MASS system. The agreement between the colours derived from these four
spectra is usually very good, the standard deviation being always below
$0.008$ mag for all indices, except for those involving the $H$ and $K_s$
band (where the standard deviation increases to $0.02-0.04$ mag).

\begin{figure*}
\begin{center}
\includegraphics[width=1\textwidth]{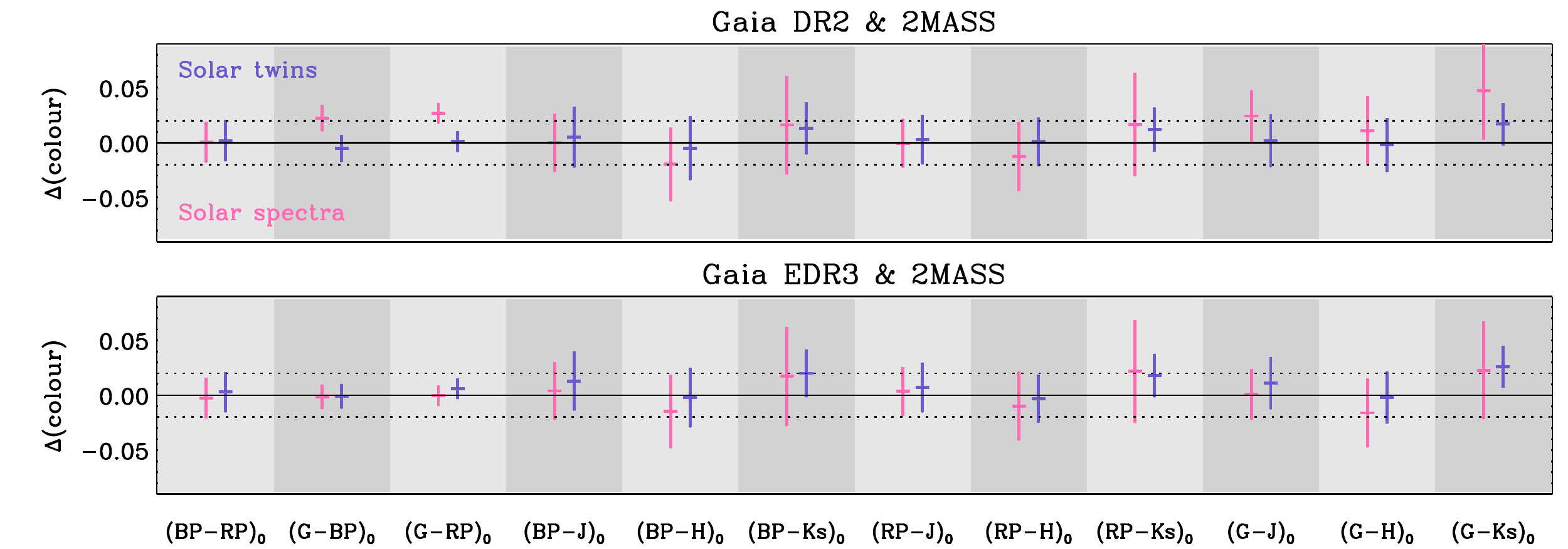}
\caption{Difference between the colours of the Sun listed in Table
  \ref{tab:sun} and those derived by averaging the colours of four absolutely
  calibrated solar reference spectra (pink). Error bars are the squared root of
  the squared sum of the uncertainties reported in Table \ref{tab:sun} and of
  the standard deviation of the colours derived from our four reference
  spectra. Also shown is the difference between our colours and those inferred
  from solar twins (blue). Again, error bars are the squared root of the
  squared sum of the uncertainties in the two dataset. Dotted lines mark
  $\pm0.02$ mag to give a better sense of the typical agreement across
  different colour indices.}\label{fig:sunco} 
\end{center}
\end{figure*}

Figure \ref{fig:sunco} shows that our inferred solar colours are in overall
excellent agreement with those obtained from solar
reference spectra and solar twins. We use the same solar twins of Table
\ref{tab:check}, which have an average spectroscopic $\teff$ centred within
couple of K from our adopted solar value (depending whether the sample from DR2
-which comprises 8 stars- or EDR3 -10 stars- is used). For the Gaia DR2
system, colours
map the effective temperature differences already discussed for Table
\ref{tab:check}. It can be appreciated how well the colours of the Sun from
different dataset agree, the difference being $\lesssim 0.02$~mag for virtually
all bands. In the EDR3 system, the agreement is particularly remarkable for
the pure Gaia colours $(BP-RP)_0$, $(G-BP)_0$ and $(G-RP)_0$, where our
temperature scale, solar twins and solar spectra all agree to better than
$0.006$~mag. This is likely indicative of how well EDR3 zero-points and
transmission curves are characterized, and how robustly solar colours can now
be derived for the Gaia system. 

\begin{table}
  \caption{Solar colours.}\label{tab:sun}
\begin{tabular}{crr}
\hline
 colour      &  Gaia DR2 - 2MASS  & Gaia EDR3 - 2MASS   \\
\hline
$(BP-RP)_0$  &  $0.823 \pm 0.018$ & $ 0.815 \pm 0.018$  \\ 
$(G-BP)_0$   & $-0.354 \pm 0.012$ & $-0.322 \pm 0.011$  \\ 
$(G-RP)_0$   &  $0.465 \pm 0.009$ & $ 0.489 \pm 0.009$  \\
$(BP-J)_0$   &  $1.372 \pm 0.025$ & $ 1.350 \pm 0.025$  \\
$(BP-H)_0$   &  $1.683 \pm 0.025$ & $ 1.660 \pm 0.024$  \\ 
$(BP-K_s)_0$ &  $1.731 \pm 0.019$ & $ 1.712 \pm 0.018$  \\ 
$(RP-J)_0$   &  $0.549 \pm 0.021$ & $ 0.538 \pm 0.021$  \\ 
$(RP-H)_0$   &  $0.852 \pm 0.020$ & $ 0.843 \pm 0.020$  \\ 
$(RP-K_s)_0$ &  $0.907 \pm 0.018$ & $ 0.895 \pm 0.018$  \\ 
$(G-J)_0$    &  $1.016 \pm 0.022$ & $ 1.030 \pm 0.022$  \\ 
$(G-H)_0$    &  $1.321 \pm 0.021$ & $ 1.338 \pm 0.021$  \\ 
$(G-K_s)_0$  &  $1.368 \pm 0.016$ & $ 1.383 \pm 0.016$  \\ 
\hline
\end{tabular}
\begin{minipage}{0.45\textwidth}
  For the Gaia DR2 system, the values provided here supersede those in \cite{cv18}. The solar absolute magnitude of the averaged flux calibrated spectra is $M_{G,\rm{DR2}}=4.675 \pm 0.006$ and $M_{G,\rm{EDR3}}=4.665 \pm 0.006$.
\end{minipage}
\end{table}
    
\newpage
\noindent \rule{8.5cm}{1pt}

\noindent
$^{1}$Research School of Astronomy and Astrophysics, The Australian National University, Canberra, ACT 2611, Australia\\
$^{2}$ARC Centre of Excellence for All Sky Astrophysics in 3 Dimensions (ASTRO 3D), Australia\\
$^{3}$Centre for Astrophysics and Supercomputing, Swinburne University of Technology, Melbourne, VIC 3122, Australia\\ 
$^{4}$Centre for Astrophysics, University of Southern Queensland, Toowoomba, QLD 4350, Australia\\
$^{5}$Max Planck Institute for Astrophysics, Karl-Schwarzschild-Str.~1, D-85748 Garching, Germany\\
$^{6}$Sydney Institute for Astronomy, School of Physics, A28, The University of Sydney, NSW 2006, Australia\\
$^{7}$School of Physics, UNSW, Sydney, NSW 2052, Australia\\
$^{8}$Institute for Advanced Study, Princeton, NJ 08540, USA\\
$^{9}$Department of Astrophysical Sciences, Princeton University, Princeton, NJ 08544, USA\\
$^{10}$Observatories of the Carnegie Institution of Washington, 813 Santa Barbara Street, Pasadena, CA 91101, USA\\
$^{11}$Monash Centre for Astrophysics, Monash University, Australia\\
$^{12}$School of Physics and Astronomy, Monash University, Australia\\
$^{13}$Australian Astronomical Optics, Faculty of Science and Engineering, Macquarie University, Macquarie Park, NSW 2113, Australia\\
$^{14}$Macquarie University Research Centre for Astronomy, Astrophysics \& Astrophotonics, Sydney, NSW 2109, Australia\\
$^{15}$Istituto Nazionale di Astrofisica, Osservatorio Astronomico di Padova, vicolo dell'Osservatorio 5, 35122, Padova, Italy\\
$^{16}$Faculty of Mathematics and Physics, University of Ljubljana, Jadranska 19, 1000 Ljubljana, Slovenia\\
$^{17}$Department of Astronomy, Stockholm University, AlbaNova University Centre, SE-106 91 Stockholm, Sweden\\
$^{18}$Department of Physics and Astronomy, Macquarie University, Sydney, NSW 2109, Australia\\

\end{document}